\documentstyle[12pt]{article}
\makeatletter 
 \renewcommand{\thesection}{%
   \Roman{section}}
 \renewcommand{\theequation}{% 
   \arabic{section}.\arabic{equation}}
 \@addtoreset{equation}{section}
\makeatother

\begin{document}

\begin{titlepage}
\null

\begin{flushright}
hep-th/9604150 \\
KOBE-TH-96-01 \\
April 1996
\end{flushright}

\begin{center}
  {\Large\bf Strong Coupling Quantum Gravity and \par}
  {\Large\bf Physics beyond the Planck Scale \par}
  \vspace{1.5cm}
  \baselineskip=7mm

  {\large  Kayoko Maeda\footnote{
  E-mail address: maeda@hetsun1.phys.kobe-u.ac.jp}
   and Makoto Sakamoto\footnote{
  E-mail address: sakamoto@hetsun1.phys.kobe-u.ac.jp} \par}
  \vspace{5mm}
   {\sl Department of Physics, Kobe University\\
        Rokkodai, Nada, Kobe 657, Japan\par}

\vspace{4cm}
{\large\bf Abstract}
\end{center}
\par

We propose a renormalization prescription for 
the Wheeler-DeWitt equation 
of (3+1)-dimensional Einstein gravity 
and also propose a strong coupling expansion as
an approximation scheme to probe quantum 
geometry at length scales much smaller than
the Planck length. We solve the Wheeler-DeWitt 
equation to the second order in the expansion 
in a class of local solutions 
and discuss problems arising in our approach.

\begin{center}
\vspace{2cm}
{\it To be published in Physical Review D}
\end{center}

\end{titlepage}
\setcounter{footnote}{0}
\baselineskip=7mm

\setcounter{footnote}{2}
%%%%%%%%%%%%%%%%%%%%%%%%%%%%%%%%%%%%%%%%%%%%%%%%%%%%%%
\section{INTRODUCTION}

Nonperturbative effects of quantum gravity would play 
a vital role 
in the physics at the Planck scale and drastically change 
the concept of spacetime below the Planck length.
The perturbative non-renormalizability of quantum 
gravity\cite{nonrenormalizability} is probably a sign that 
nonperturbative approaches to quantum gravity are essential
to study spacetime structure at short distances 
below the Planck length.

Although we have not yet understood the physics at 
and beyond the Planck 
scale, there have been frequent suggestions that 
the concept of spacetime 
loses its meaning below the Planck length. In fact, 
many different 
approaches to quantum gravity have led to 
the conclusion of the existence 
of a minimum length\footnote{For a recent review, 
see Ref.\cite{minimum length}.}.
The concept of the minimum length is very suggestive. 
Since it may imply
discrete nature of spacetime in 
quantum gravity\cite{discrete spacetime}, 
the number of dynamical degrees of freedom will be 
much smaller than what one naively expects. 
Witten has proposed an interesting idea that 
the physics at and beyond the
Planck scale is described by a topological 
quantum field theory with
a finite dimensional Hilbert space\cite{TQFT}.  
It has also been argued by 't Hooft that the finiteness 
of entropy and information in a black hole is evidence 
for the discreteness of 
spacetime and that the number of degrees of 
freedom is given by 
the area of the event horizon in Planck units.
This has led 't Hooft\cite{dimensional reduction} 
and Susskind\cite{hologram}
to make the holographic hypothesis 
that physical states are described 
by a quantum field theory on the surface of the black hole, 
and a realization of the idea has been discussed by Smolin 
in topological quantum field theory\cite{smolin}.

A canonical quantization of gravity is one of basic 
approaches to quantum gravity. 
A major advance has been developed in the canonical 
quantum gravity proposed by 
Ashtekar\cite{Ashtekar}.
In terms of Ashtekar's new variables, a large class of 
solutions to 
the Hamiltonian constraint has been constructed in 
the loop representation\cite{loop}.
A remarkable observation is that the area 
and the volume operators 
have discrete spectrum, i.e., they are quantized 
in Planck units\cite{area operator, volume}. 
 
In this paper, we will take another approach to 
the canonical quantum gravity 
based on the Wheeler-DeWitt equation\cite{WDW}
and investigate spacetime structure beyond the Planck scale. 
Main technical obstacles to prevent the study of 
the short distance physics are 
that WKB approximation will 
not be applicable at short distances 
and that the Wheeler-DeWitt equation is 
ill-defined without regularization.
In this paper, we propose an approximation scheme 
which will be well-suited to 
probe quantum geometry at length scales much smaller 
than the Planck length, 
and also propose a renormalization prescription 
to make the Wheeler-DeWitt 
equation finite. Some of our results have been 
reported in Ref.\cite{analysis}. 
We shall give a full detail of Ref.\cite{analysis} 
and results to the next 
order approximation. We shall also discuss various 
problems arising in our
approach.

This paper is organized as follows: 
In Sec. II, we explain our renormalization prescription 
for the Wheeler-DeWitt 
equation. 
In Sec. III, we check consistency of 
the constraints with our
renormalization prescription. 
In Sec. IV, we explain that the strong coupling 
expansion is well-suited 
to study the physics beyond the Planck scale. 
In Sec. V, we look for solutions of the Wheeler-DeWitt 
equation to the first order
in the strong coupling expansion. 
In Sec. VI, we solve the Wheeler-DeWitt equation 
to the second order in the 
strong coupling expansion and discuss a problem 
arising in higher order calculations.
Sec. VII is devoted to conclusion. 
In three Appendices A, B and C, a detail of 
computations of results in 
Sec. II and III are given.
   
%%%%%%%%%%%%%%%%%%%%%%%%%%%%%%%%%%%%%%%%%%%%%%%%%%%%%%%%%%%%%
\section{REGULARIZATION PRESCRIPTION} 

The (unregulated) Wheeler-DeWitt equation without matter is
\begin{equation}
\biggl[\, G_{ijkl}(x) 
    \frac{\delta}{\delta h_{kl}(x)}
    \frac{\delta}{\delta h_{ij}(x)}
- \frac{1}{(16 \pi G)^2}\,\sqrt{h(x)}
     \Bigl(\!~ R(x) + 2 \Lambda \Bigr) \biggr]\, 
      \Psi[h] = 0\   ,
\label{201}
\end{equation}
where $G$ is the Newton constant and
$G_{ijkl}$ is the metric on superspace
\begin{equation}
G_{ijkl}= \frac{1}{2 \sqrt{h}} (h_{ik} h_{jl} + h_{il} h_{jk}
            - h_{ij} h_{kl})\ .
\label{202}
\end{equation}
The $R(x)$ denotes the scalar curvature constructed from 
the three-metric $h_{ij}$ and $\Lambda$ is the cosmological 
constant.
The Wheeler-DeWitt equation needs regularization
because it contains a
product of two functional derivatives at the same spatial point,
\begin{equation}
\Delta(x) \equiv G_{ijkl} (x)
  \frac{\delta}{\delta h_{kl}(x)}
  \frac{\delta}{\delta h_{ij}(x)}\ .
\label{203}
\end{equation}
For example, $\Delta(x)$ acting on $R(y)$ is proportional to 
$(\delta(x,y))^2$, which is meaningless.
To make (\ref{201}) well defined, we want to replace $\Delta(x)$ 
by a renormalized operator $\Delta_{R}(x)$. 
We will require $\Delta_{R}(x)$ to be a finite operator
preserving three-dimensional general coordinate invariance 
and also to be consistent with the constraints 
which are the generators of the symmetry of the theory.

Recently Mansfield proposed a renormalization scheme
to solve the Schr\"odinger equation for 
Yang-Mills theory in the strong coupling 
expansion\cite{Mansfield}.
We shall generalize
the renormalization procedure developed  by Mansfield 
to the Wheeler-DeWitt equation.
Our renormalization prescription consists of two steps: 
The first step is to regularize the operator 
(\ref{203}) by point-splitting the functional derivatives.
The second step is to remove a cutoff by using analytic 
continuation and to extract finite quantities.

\ \ 

\noindent
{\large\bf A. Regularization scheme}

The first step to construct $\Delta_R(x)$ is to \lq\lq point-split" 
the functional derivatives by use of a heat kernel\cite{Mansfield}.
We replace $\Delta(x)$ by the following differential operator:
\begin{equation}
\Delta(x;t) \equiv 
   \int d^3x' K_{i'j'kl}(x',x;t)   
      \frac{\delta}{\delta h_{kl}(x)}
      \frac{\delta}{\delta h_{i'j'}(x')}\ , 
\label{204}
\end{equation}
where $K_{i'j'kl}(x',x;t)$ is a bi-tensor 
at both $x'$ and $x$ and satisfies the heat equation,
\begin{equation}
- \frac{\partial}{\partial t}  K_{i'j'kl}(x',x;t) =
- ({\nabla'}\!\!_{p} {\nabla'}^{p}  + \xi R(x') ) 
K_{i'j'kl}(x',x;t)\ ,
\label{205}
\end{equation}
with the initial condition
\begin{equation}
 \lim_{t \rightarrow 0}  K_{i'j'kl}(x',x;t) 
 = G_{i'j'kl}(x) \delta(x',x)\ .
\label{206}
\end{equation}
Here, ${\nabla'}\!\!_{p}$ and $\delta(x',x)$ denote the covariant
derivative with respect to $x'$ and 
the three-dimensional $\delta$ function,
respectively, and $\xi$ is an arbitrary constant. 
Taking $t$ small but nonzero in (\ref{204}) gives a
regulated operator of $\Delta(x)$. 
(In the naive limit $t \rightarrow 0$, $\Delta(x;t)$ is 
reduced to $\Delta(x)$.) 
In Ref.\cite{analysis}, the heat kernel $K_{i'j'kl}$ 
has been assumed to satisfy
(\ref{205})  with $\xi = 0$. We will see in Sec. VI~ 
that the term $\xi R$ in (\ref{205}) plays 
an important role in finding solutions of 
the Wheeler-DeWitt equation to the second 
order in the strong coupling expansion.

We should make a few comments on 
the regulated operator (\ref{204}). 
We have 
chosen the factor ordering written 
in (\ref{204}). Other choices of factor 
ordering will lead to different values 
of numerical constants of our results
but will not change the qualitative features. 
Although the heat 
equation (\ref{205}) with the initial 
condition (\ref{206})
has been chosen to be consistent with 
the requirement of preserving 
three-dimensional general coordinate invariance, 
it is not a unique choice to 
satisfy the requirement. For example, one might 
add $m^2$ or $\eta R^2(x')$
to ${\nabla'}\!\!_{p} {\nabla'}^{p}  + \xi R(x')$ 
in (\ref{205}).
Thus, the wave functional $\Psi[h]$ depends on the choice of
the heat equation and more generally on the choice 
of renormalization prescriptions. 
A renormalization prescription can make the form of 
a solution very simple but 
other renormalization prescriptions may make 
it complicated\footnote{ 
A similar situation occurs in gauge theories. 
A wave functional for 
non-abelian gauge theories quite depends on 
the choice of gauge fixings.
Wilson and co-workers have recently proposed that the light-cone
gauge is well-suited to study nonperturbative dynamics of QCD 
at low energies\cite{light-front QCD} }.
This fact does not , however, necessarily mean 
that physical quantities would
depend on the choice of the heat equation, 
because the wave functional itself 
is not a physical observable. We thus hope 
that physical quantities are
independent of any choice of the heat equation. 
We will not justify it in this paper. 

The heat equation (\ref{205}) can be solved 
by the standard technique\cite{Schwinger-DeWitt}.
For our purpose we need to know only $K_{i'j'kl}(x',x;t)$ 
and some of its 
covariant derivatives in the limit $x' \rightarrow x$. 
We will give their explicit forms in Appendix A. 
Let ${\cal O}$ be three-dimensional integrals of 
local functions of $h_{ij}$. 
The action of $\Delta(x;t)$ on ${\cal O}$ 
will give an expansion in 
powers of $t$.
These powers of $t$ may be determined from 
general coordinate invariance
and dimensional analysis ( $t$ and $\Delta(x;t)$ have mass dimension
$-2$ and $6$, respectively). For example, we have  
\begin{eqnarray}
\lefteqn{\Delta(x;t) \int d^3y \sqrt{h} =
         \frac{\sqrt{h(x)}}{(4 \pi)^{3/2}}
         \biggl\{ \frac{\alpha_1}{t^{3/2}} + O(t^{-1/2}) \biggr\}, }
\label{207}  \\
\lefteqn{\Delta(x;t) \int d^3y \sqrt{h}~ R
    =   \frac{\sqrt{h(x)}}{(4 \pi)^{3/2}}
        \biggl\{  \frac{\beta_1}{t^{5/2}} 
                + \frac{\beta_2}{t^{3/2}} R(x) 
                + O(t^{-1/2}) \biggr\}, }
\label{208} \\
\lefteqn{\Delta(x;t) \int d^3y \sqrt{h}~ R^2  
   =   \frac{\sqrt{h(x)}}{(4 \pi)^{3/2}}
        \biggl\{  \frac{\gamma_1}{t^{7/2}} 
                + \frac{\gamma_2}{t^{5/2}} R(x)  } \nonumber \\
  &&          + \frac{1}{t^{3/2}} \Bigl(  \gamma_3 ~R^2(x) 
                +  \gamma_4  ~R_{ij}(x)\! ~R^{ij}(x)
                + \gamma_5 \nabla\!_i \nabla^i\! R(x)  \Bigr)  
        + O(t^{-1/2}) \biggr\}, 
\label{209} \\
\lefteqn{\Delta(x;t) \int d^3y \sqrt{h}~ R_{ij}R^{ij}
   =  \frac{\sqrt{h(x)}}{(4 \pi)^{3/2}}
        \biggl\{  \frac{{\gamma'}_1}{t^{7/2}} 
                + \frac{{\gamma'}_2}{t^{5/2}} ~ R(x) } \nonumber\\
   & &          + \frac{1}{t^{3/2}} \Bigl(  {\gamma'}_3  R^2(x) 
                + {\gamma'}_4  ~R_{ij}(x) ~ R^{ij}(x)
                + {\gamma'}_5 \nabla\!_i \nabla^i\! ~R(x)  \Bigr)  
             + O(t^{-1/2}) \biggr\}.
\label{210}
\end{eqnarray}
The first few coefficients are given by 
\begin{eqnarray}
&&\alpha_1 = - \frac{21}{ 8} ~~~~ , ~~~~
  \beta_1 =   \frac{ 3}{ 2} ~~~~ , ~~~~  
  \beta_2 = - \frac{11}{24} + \frac{3}{2}~ \xi ~~~~ ,
\label{211} \\
&&\gamma_1 = 0 \ , ~~
  \gamma_2 = - 3 \ , ~~
  \gamma_3 =  \displaystyle{\frac{11}{8} } - 3 ~ \xi \ , ~~
  \gamma_4 = \frac{31}{6} \ , ~~
  \gamma_5 = -1 \ , \nonumber\\
&&{\gamma'}_1 = \displaystyle{\frac{15}{4}} \ , ~~
  {\gamma'}_2 = \displaystyle{
  - \frac{13}{8}+\frac{15}{4} ~ \xi } \ , ~~
  {\gamma'}_3 = \displaystyle{
  -\frac{241}{480}-\frac{13}{8} \xi +\frac{15}{8} ~ \xi^2 }\ ,
  \nonumber\\
&&{\gamma'}_4 = \displaystyle{\frac{97}{20} }\ , ~~
  {\gamma'}_5 = \displaystyle{-\frac{11}{30}-\frac{25}{24} ~ \xi }\ .
\label{212}
\end{eqnarray}
The coefficients $\alpha_1$, $\beta_1$ and $\beta_2$ agree 
with those in Ref.\cite{analysis}
with $\xi = 0$. A detail of the computations to get the coefficients 
(\ref{211}) and (\ref{212}) will be found in Appendix B. 
The above results will 
be used later in the discussion of finding solutions 
to the Wheeler-DeWitt 
equation in the strong coupling expansion. 

\ \ 

\noindent
{\large\bf B. Analytic continuation }

The second step  of our renormalization prescription is
to extract a finite part from $ \Delta(x;t=0) {\cal O}$.
Note that we cannot simply 
take the limit $t \rightarrow 0$ in $\Delta(x;t) {\cal O}$ 
because of the presence of inverse powers of
$t$ (see (\ref{207}) to (\ref{210}) ). 
We will here define $\Delta_R(x) {\cal O}$ from 
$\Delta(x;t) {\cal O}$ by analytic continuation 
so that $\Delta_R(x) {\cal O}$ is identical to 
$\Delta(x;t=0) {\cal O}$ (and hence $\Delta(x) {\cal O}$)
if $\Delta(x;t=0) {\cal O}$ is finite. 
Our definition of $\Delta_R(x) {\cal O}$ is given by
\begin{equation}
\Delta_{R}(x) {\cal O} ~\equiv~ 
  \lim_{s \rightarrow +0} ~
  s \int^{\infty}_{0} d\varepsilon\, 
      \varepsilon^{s-1} \phi(\varepsilon) 
      \Delta(x;t=\varepsilon^2)\, {\cal O}\ ~~~.
\label{213}
\end{equation}
It is easy to see that $\Delta_R(x) {\cal O}$ 
is equal to $\Delta(x;t=0) {\cal O}$ if 
$\Delta(x;t=0) {\cal O}$ is finite 
as long as a differentiable function $\phi(\varepsilon)$
rapidly decreases to zero at infinity with $\phi(0)=1$,
\begin{equation}
\left\{
\begin{array}{lcll}
\phi ~ (\infty) &=& 0 & ,  \vspace{3mm}\\ 
\phi ~ (0) &=&1 & .
\end{array} \right.
\label{214}
\end{equation} 
By analytic continuation, 
we can give a meaning to the right hand side
of (\ref{213}) even if 
$\Delta(x;\varepsilon^2) {\cal O}$ diverges 
at the origin like $\varepsilon^{-n}$ with integer $n$:
The integral (\ref{213}) exists for $ s > n$
( provided 
$\phi(\varepsilon) \Delta(x; \varepsilon^2) {\cal O}$ 
has no other divergences )
so that we can analytically continue $s$ from  $ s > n$ 
to small values and take the limit 
$s \rightarrow 0$ to obtain a finite result\footnote{ 
For the detail, see Ref \cite{Mansfield}.}.
For example, our definition of $\Delta_R(x){\cal O}$ for 
$\Delta(x;\varepsilon^2) {\cal O} = f(x) {\varepsilon}^{-n}$ 
gives  
\begin{equation}
\Delta_R(x) {\cal O} = \left\{
\begin{array}{ll}
0  & {\rm for} ~~~~  n = -1,-2,-3,\cdots ~~ ,\vspace{3mm}\\
\displaystyle{
\frac{1}{n!}\frac{d^n \phi(0)}{d \varepsilon^n} f(x) } 
& {\rm for} ~~~~ n = 0,1,2,\cdots ~~ .
\end{array} \right.
\label{215}
\end{equation}
In this stage, the function 
$\phi(\varepsilon)$ is arbitrary as long as it 
satisfies the conditions (\ref{214}).
As we will see in the next section, $\phi(\varepsilon)$ 
has to be subject to a further condition 
$\frac{d\phi(0)}{d\varepsilon}=0$.
For definiteness, we may choose $\phi(\varepsilon)$ 
to be of the form 
\begin{equation}
\phi(\varepsilon) 
 ~=~ (1 + \mu \varepsilon)~ {\rm e}^{- \mu \varepsilon} 
\label{216}
\end{equation}
which satisfies all requirements, though it is not necessary
to choose $\phi(\varepsilon)$ as above. 
Here, $\mu$ is an arbitrary mass parameter. 
Then, the equations (\ref{207}) to (\ref{210}) are now 
replaced by 
\begin{eqnarray}
\lefteqn{  \Delta_{R}(x) \int d^3y \sqrt{h} =
      \frac{\sqrt{h(x)}}{(4 \pi)^{3/2}}
      \biggl\{ \frac{\phi^{(3)}(0)}{3!} \alpha_1 \biggr\} \ ,  }
\label{217} \\
\lefteqn{ \Delta_{R}(x) \int d^3y \sqrt{h} ~ R =
     \frac{\sqrt{h(x)}}{(4 \pi)^{3/2}} 
     \biggl\{  \frac{\phi^{(5)}(0)}{5!} \beta_1
           + \frac{\phi^{(3)}(0)}{3!} \beta_2  ~ R(x)  
           \biggr\}\ , }
\label{218} \\
\lefteqn{ \Delta_{R}(x) \int d^3y \sqrt{h} ~ R^2 =
     \frac{\sqrt{h(x)}}{(4 \pi)^{3/2}} 
     \biggl\{  \frac{\phi^{(7)}(0)}{7!} \gamma_1 
             + \frac{\phi^{(5)}(0)}{5!} \gamma_2 ~ R(x) 
             } \nonumber\\ 
&& \hspace{15mm}+ \frac{\phi^{(3)}(0)}{3!} 
                \Bigl( \gamma_3 ~ R^2(x)  
                    +  \gamma_4\! ~ R_{ij}(x)\! ~ R^{ij}(x)
                    +  \gamma_5 \nabla\!_i \nabla^i\! ~ R(x)
                \Bigr)\biggr\}\  , 
\label{219} \\
\lefteqn{ 
     \Delta_{R}(x) \int d^3y \sqrt{h} ~ R_{ij}\! ~ R^{ij} =
     \frac{\sqrt{h(x)}}{(4 \pi)^{3/2}} 
     \biggl\{  \frac{\phi^{(7)}(0)}{7!} {\gamma'}_1 
             + \frac{\phi^{(5)}(0)}{5!} {\gamma'}_2 R(x) 
                } \nonumber\\ 
&& \hspace{15mm}+  \frac{\phi^{(3)}(0)}{3!}
                \Bigl(  {\gamma'}_3 ~ R^2(x) 
                      + {\gamma'}_4\! ~ R_{ij}(x)\! ~ R^{ij}(x)
                      + {\gamma'}_5 \nabla\!_i \nabla^i\! ~ R(x)
                \Bigr)\biggr\}\  ,   
\label{220}
\end{eqnarray}
where
\begin{equation}
\phi^{(n)}(0)
 \equiv 
  \displaystyle{
  \frac{d^n \phi(0)}{d \varepsilon^n} } 
  ~=~ (-1)^{n-1}(n-1) \mu^n ~~~.
\label{221}
\end{equation}
The results (\ref{217}) to (\ref{220}) depend on the 
arbitrary mass parameter $\mu$. This is an inevitable consequence
of isolating finite quantities from divergent ones. ( For 
instance, in the dimensional regularization\cite{dimred}
an arbitrary mass parameter is introduced for coupling 
constants to have proper dimensions.) 
Physical observables must be independent of this arbitrary  
parameter $\mu$. This leads to a renormalization group equation 
so that coupling \lq\lq constants" should be regarded as
functions of $\mu$. This is the basic problem of renormalization. 
We shall return to this point in Sec. V and explicitly show 
that the $\mu$-dependence can be absorbed into the redefinition 
of the Newton constant $G$ and the cosmological constant $\Lambda$ 
to the first order in the strong coupling expansion.

%%%%%%%%%%%%%%%%%%%%%%%%%%%%%%%%%%%%%%%%%%%%%%%%%%%%%%%%%%%%%%%
 \section{CONSISTENCY OF CONSTRAINTS}

We have chosen the renormalization prescription 
to preserve three-dimensional 
general coordinate invariance but this is not enough to preserve 
the whole symmetry of the theory at the quantum level. 
We have to check that our renormalization
prescription would be consistent with the constraints 
which are the generators of the symmetry. 
Consistency of the constraints 
requires that commutators of the constraints 
lead to no new constraints. 

The constraints consist of the 
momentum constraint $ {\cal H}_i(x)$ and 
the Hamiltonian constraint ${\cal H}(x)$: 
\begin{eqnarray} 
{\cal H}_i (x) &\equiv& - 2 h_{ij} \nabla_k \pi^{jk}(x) \ ,  
\label{301} \\
{\cal H}(x) &\equiv&  16 \pi G \, 
G_{ijkl}(x) \pi^{kl}(x) \pi^{ij}(x) 
   + \frac{1}{16\pi G}\sqrt{h(x)} 
     \Bigl( R(x) + 2 \Lambda \Bigr) \ ,    
\label{302}
\end{eqnarray}
where $\pi^{ij}(x) = - i \frac{\delta}{\delta h_{ij}(x)}$ is
the momentum operator. We will take the factor ordering 
such that $\pi$'s stand to the right and $h$'s stand to the left, 
as written above. The momentum constraints ${\cal H}_i$'s are the
generators of three-dimensional general coordinate transformations. 
Since our renormalization prescription preserves three-dimensional 
general coordinate invariance, no anomalous terms may appear in 
commutators with the momentum constraints.

There remains to be considered only the commutator of 
the Hamiltonian constraints. In our renormalization prescription, 
the Hamiltonian constraint (\ref{302}) should be replaced by 
\begin{equation}
{\cal H}_R(x) \equiv - 16 \pi G \, \Delta_{R}(x) 
                   + \frac{1}{16 \pi G}\sqrt{h(x)} 
                    \Bigl( R(x) + 2 \Lambda \Bigr)\ . 
\label{303}
\end{equation}
 The commutator $ [{\cal H}_R(x) ,{\cal H}_R(y) ]$, more correctly, 
$ [ \int d^3x \eta_1(x) {\cal H}_R(x) , 
\int d^3y \eta_2(y) {\cal H}_R(y) ]$ 
will have the form 
\begin{eqnarray}
\lefteqn{
\Bigl[~~ \int d^3x~ \eta_1(x) ~{\cal H}_R(x) ~~, 
~~ \int d^3y~ \eta_2(y)~ {\cal H}_R(y) ~~ \Bigr] } \nonumber \\
&& = i \int d^3x 
 ~\Bigl( \eta_1(x) ( \nabla\!_i\, \eta_2(x) ) 
 - (\nabla\!_i\, \eta_1 (x)) \eta_2 (x) \Bigr) {\cal H}^i(x)
  ~+~ \Delta \Gamma  \ , 
\label{304}
\end{eqnarray}
for arbitrary scalar functions $\eta_1$ and $\eta_2$. 
An anomalous term $\Delta \Gamma$ could appear from 
the commutators of $\Delta_{R}$ 
and $\sqrt{h} ~ R$. 
It follows from dimensional analysis and 
the antisymmetry under the exchange 
of $\eta_1$ and $\eta_2$ that $\Delta \Gamma$ is expected to be of 
the form\footnote{The $\phi^{(n)}(0)$ 
has mass dimension $n$.}\cite{analysis}.
\begin{equation}
\Delta \Gamma \equiv 
  a~ \phi^{(1)}(0) \int d^3x \sqrt{h(x)} 
 ~\Bigl(  \eta_1(x) ( \nabla\!_i\, \eta_2(x) ) 
  - (\nabla\!_i\, \eta_1 (x)) \eta_2 (x) \Bigr)~
   \nabla^i\! R(x) \ . 
\label{305}
\end{equation}
In Appendix C, we will compute the commutator (\ref{304}) and find 
$a = - ( \frac{1}{24} + \xi) /(4 \pi )^{3/2} $.
We therefore take 
\begin{equation}
\phi^{(1)}(0) = 0
\label{306}
\end{equation}
for consistency of the constraints with 
our renormalization prescription, 
as claimed in the previous section.

We should make a comment on the anomaly free condition 
$\Delta \Gamma = 0$.
One may take $\xi = - \frac{1}{24}$, instead of 
the condition (\ref{306}),
to have $\Delta \Gamma = 0$. As mentioned in the previous section, 
the heat equation (\ref{205}) is not a unique choice to satisfy our 
requirements and there is a great deal of 
freedom to modify the heat equation. 
We can, however, show that $\Delta \Gamma$ 
is proportional to $\phi^{(1)}(0)$,
irrespective of what heat equation we choose. 
Hence, the choice (\ref{306}) 
seems more universal than the choice $\xi = -\frac{1}{24}$. 
%because (\ref{306}) leads to the anomaly free condition 
%$\Delta \Gamma = 0 $ for any choice of the heat equation. 
We will see later that the free parameter $\xi$ is used in 
finding solutions to the second order 
in the strong coupling expansion.

It should be noticed that the above analysis of consistency of the 
constraints is incomplete. Although we have found 
an anomalous term (\ref{305}), which leads to the condition 
(\ref{306}), it does not imply that no other anomalous terms 
would appear. This is because we have not exactly computed 
the commutators of the constraints due to lack of our knowledge 
of the exact expressions of 
$\frac{\delta K_{ijkl}(x,y;t)}{\delta h_{mn}(z)} $
and
$\frac{\delta^2 K_{ijkl}(x,y;t)}
      {\delta h_{mn}(z) \delta h_{pq}(z') } $
for arbitrary $x, y, z$ and $z'$. 
The exact computations of the commutators, 
$ [{\cal H}_i(x) ,{\cal H}_R(y) ]$ 
and
$ [{\cal H}_R(x) ,{\cal H}_R(y) ]$, 
remain to be performed.
If other anomalous terms were shown to appear, we have to impose 
further conditions besides (\ref{306}) to be anomaly free. 

%%%%%%%%%%%%%%%%%%%%%%%%%%%%%%%%%%%%%%%%%%%%%%%%%%%%
\section{STRONG COUPLING EXPANSION}

As discussed in the previous sections, 
we have the renormalized Wheeler-DeWitt equation, 
\begin{equation}
   \biggl[\, \Delta_{R}(x)
   - \frac{1}{(16 \pi G)^2} \sqrt{h(x)}
     \Bigl(\!~ R(x) + 2 \Lambda \Bigr) \biggr]\, \Psi[h] = 0\  ,
\label{401}
\end{equation}
with the condition (\ref{306}). It is not easy to 
solve the full Wheeler-DeWitt 
equation exactly. Since we are interested 
in short distance behavior of 
the Wheeler-DeWitt equation, we do not probably 
need to solve (\ref{401}) 
exactly. The Planck length is given by $l_p = (\hbar G/c^3)^{1/2} $
so that the strong coupling limit, i.e., $G \rightarrow \infty$, 
will be well-suited to probe quantum geometry at length scales much 
{\it smaller} than the Planck length\cite{strong}. 
We will discuss this point in some detail later. 
Although strong coupling 
quantum gravity has been studied before, 
their studies are not satisfactory 
to clarify quantum geometry at short distances
 because little attention has 
been paid to the regularization of the Wheeler-DeWitt equation. 

To solve the equation (\ref{401}) with the condition (\ref{306}), 
we attempt an expansion of the wave functional of the universe in 
inverse powers of the Newton constant $G$. 
We will first rewrite the 
Wheeler-DeWitt equation (\ref{401}) as 
\begin{equation}
\Delta_{R}(x)\, S[h] 
  - G_{ijkl}(x) \frac{\delta S[h]}{\delta h_{kl}(x)} 
    \frac{\delta S[h]}{\delta h_{ij}(x)} 
  + \frac{1}{(16\pi G)^2} \sqrt{h(x)} 
      \Bigl( R(x) + 2 \Lambda \Bigr) = 0\  ,
\label{402}
\end{equation}
where 
\begin{equation}
\Psi[h] \equiv {\rm exp} \{ -S[h] \} \ .
\label{403}
\end{equation}
We then assume that the functional $S[h]$ has the form
\begin{equation}
S[h] = \sum^{\infty}_{n= 0} 
\Bigl( \frac{1}{16 \pi G} \Bigr)^{2n} S_n[h]  \ .
\label{404}
\end{equation}
Substituting (\ref{404}) into (\ref{402}), we have , 
according to the inverse powers of $G$, 
\begin{eqnarray}
&&\Delta_{R}(x)\, S_0[h]
  - G_{ijkl}(x) \frac{\delta S_0[h]}{\delta h_{kl}(x)} 
    \frac{\delta S_0[h]}{\delta h_{ij}(x)}  = 0\  ,
\label{405} \\
\nonumber \\
&&\Delta_{R}(x)\, S_1[h] 
  - 2~ G_{ijkl}(x) \frac{\delta S_0[h]}{\delta h_{kl}(x)} 
    \frac{\delta S_1[h]}{\delta h_{ij}(x)} 
  + \sqrt{h(x)}  \Bigl(\! ~ R(x) + 2 \Lambda \Bigr) = 0\  ,
\label{406} \\
\nonumber \\
&&\Delta_{R}(x)\, S_n[h] 
  -  G_{ijkl}(x)  \sum^{n}_{m=0}
    \frac{\delta S_m[h]}{\delta h_{kl}(x)} 
    \frac{\delta S_{n-m}[h]}{\delta h_{ij}(x)}   = 0 ~~,
    ~~~~ n=2,3,4 \cdots.
\label{407}
\end{eqnarray} 
To solve the Wheeler-DeWitt equation 
in the strong coupling expansion, 
we adopt an ansatz of locality\cite{Mansfield}. 
The functional $S_n[h]$ 
is assumed to be a sum of integrals of local functions of $h_{ij}$. 
There seem no obvious reasons to restrict our attention to a 
class of local solutions because non-locality is all over in 
this theory. A main reason why we adopt the locality ansatz is
technical difficulties to find general solutions. Although 
physical significance of local solutions are quite unclear, 
it may worth while studying them to see how our formulation works.

Under the locality ansatz we find 
two solutions to the zeroth order equation 
(\ref{405}), i.e., 
\begin{equation}
S_0[h] = 0 \ ,
\label{408}
\end{equation}
and
\begin{equation}
S_0 [h] =  
\frac{7 \mu^3}{3 (4 \pi)^{3/2}} \int d^3x \sqrt{h(x)} \ .
\label{409}
\end{equation}
(To see this, we can use the relation 
 (\ref{217}) and (\ref{221}).)

We would like to make two comments on the expansion (\ref{404}). 
In Ref.\cite{analysis}, the expansion in (\ref{404}) has been  
assumed to begin with $n=1$, which corresponds 
to the first solution (\ref{408}).
We here found another solution, though 
it turns out that the second solution 
(\ref{409}) leads to essentially the same solution to higher orders 
as the first one. More generally, one might expand $S[h]$ as 
\begin{equation} 
\sum^{\infty}_{n=- m} 
\Bigl( \frac{1}{16 \pi G} \Bigr)^{2n} S_n[h] 
\label{410}
\end{equation}
with some positive integer $m$. To leading order, we would have 
\begin{equation}
  G_{ijkl}(x) \frac{\delta S_{-m}[h]}{\delta h_{kl}(x)} 
    \frac{\delta S_{-m}[h]}{\delta h_{ij}(x)}  = 0\  ,
\label{411}
\end{equation}
which will lead to $S_{-m} = 0$ under the locality ansatz. 
Then, $S_{-m+1}$ is found to satisfy the same equation (\ref{411}) 
with the replacement $ -m \rightarrow -m+1$.
Repeating the same step above, 
we finally conclude that $S_{-m} = 0$ 
for $m \geq 1$. The second comment concerns the powers of the 
Newton constant $G$ in the expansion (\ref{404}). 
We have assumed that the wave functional is expanded in powers of 
$G^{-2}$. We can, however, assume that the wave functional is more 
generally expanded in powers of $G^{-1}$, which will still give a 
consistent expansion of the equation (\ref{402}). In this paper, 
we will restrict our considerations 
to the expansion (\ref{404}) since 
(\ref{404}) respects a symmetry ($G \rightarrow -G $) of the 
Wheeler-DeWitt equation, 
but the generalization to a power series of 
$G^{-1}$ is straightforward.

Before closing this section, we would like 
to discuss physical meanings
of the strong coupling expansion in more detail. 
The semiclassical 
expansion corresponds to the expansion in powers of $\hbar$, while 
the strong coupling expansion (\ref{404}) 
corresponds to the expansion in 
powers of $\hbar^{-2}$ because $\hbar$ appears in the combination 
of $\hbar G$ in quantum gravity. 
Thus the wave functional in the strong 
coupling expansion is expected to describe 
quite different physics from 
the semiclassical one. 
The semiclassical approximation will be valid 
for long-wavelength gravitational fields, 
while the strong coupling 
approximation may be useful when 
the wavelength of gravitational fields 
very rapidly changes in a wavelength. 

Although we have mentioned that 
the strong coupling expansion is well-suited
to study the physics at length scales 
much smaller than the Planck length, we
want to discuss this point more precisely in our framework. 
We will here consider the first solution (\ref{408}) 
to the zeroth order, 
i.e., $S_0[h] = 0$. 
For the strong coupling expansion (\ref{404}) to be 
meaningful, the successive terms 
in the series for $S[h]$ should satisfy, 
in particular, 
\begin{equation}
\Bigl( \frac{1}{16 \pi G} \Bigr)^{2} S_1[h]  \gg 
\Bigl( \frac{1}{16 \pi G} \Bigr)^{4} S_2[h]  ~~~~ . 
\label{412}
\end{equation}
Taking $\Lambda =0$ for simplicity 
and using $\phi^{(n)}(0) \sim \mu^n$, 
we can write the first order term as 
\begin{equation}
\Bigl( \frac{1}{16 \pi G } \Bigr)^2 S_1[h] = 
 \Bigl( \frac{1}{ G\mu^2} \Bigr)^2
\int d^3x \sqrt{h(x)} \mu^3  
\biggl\{ c_1 +c_2 \frac{R(x)}{\mu^2} \biggr\} ~~~ ,
\label{413}
\end{equation} 
as we will find in the next section. 
The $c_n$'s are dimensionless constants 
of order one. The second order term will be shown, 
in Sec. VI, to take the 
form
\begin{eqnarray}
\lefteqn{ \Bigl( \frac{1}{16 \pi G} \Bigr)^4 S_2[h] 
  =   \Bigl( \frac{1}{ G\mu^2} \Bigr)^4
   \int d^3x \sqrt{h(x)} \mu^3  } \nonumber  \\
&&\hspace{10mm} \times \biggl\{ {c'}_1  + {c'}_2 \frac{R(x)}{\mu^2} 
        + \frac{1}{\mu^4} \Bigl(  {c'}_3 R^2(x)
        + {c'}_4 R_{ij}(x) R^{ij}(x) \Bigr) \biggr\}\  ,
\label{414}
\end{eqnarray}
where ${c'}_n$'s are dimensionless constants of order one. 
It follows that we can drop the second order term $S_2[h]$, 
provided that $ G\mu^2 \gg 1$ (i.e., $\mu \gg m_p$), $R \sim \mu^2$ 
and $R_{ij} R^{ij} \sim \mu ^4 $. We therefore expect that the 
wave functional to the first order term well describes the universe 
with a curvature much larger than the Planck mass or with a radius 
much smaller than the Planck length. It should be emphasized 
that the strong coupling expansion is not a derivative expansion 
because higher derivative terms 
like $ ( \frac{R}{\mu^2})^m ( m \geq 2)$ 
could appear on the right hand side of (\ref{413}) 
but it happens that 
their coefficients are zero as a solution 
to the equation (\ref{406}).

If we apply the strong coupling expansion (\ref{404}) 
to a minisuperspace 
model, we can see that the expansion 
is essentially identical to a 
small radius expansion of the universe. 
Thus we may expect that our
expansion gives a good approximation for 
a universe with a radius much 
smaller than the Planck length, 
though the minisuperspace model will 
not be applicable to this region.

%%%%%%%%%%%%%%%%%%%%%%%%%%%%%%%%%%%%%%%%%%%%%%%%%%%%%%%%%%%%%%%%
\section{THE FIRST ORDER SOLUTIONS} 

In this section, we shall look for solutions to $S_1[h]$ 
in the strong coupling
expansion. Since both zeroth order solutions (\ref{408}) 
and (\ref{409}) lead 
to essentially the same solution to $S_1[h]$, we will 
mainly consider the 
zeroth order solution (\ref{408}). 
Substituting $S_0[h] = 0$ into the first 
order equation (\ref{406}), we get 
\begin{equation}
\Delta_{R}(x)\, S_1[h] = - \sqrt{h(x)}  
\Bigl(\! ~ R(x) + 2 \Lambda \Bigr)  \  .
\label{501}
\end{equation} 
From (\ref{217}) and (\ref{218}), we find a solution 
to the first order 
as\footnote{ In Ref.\cite{Kodama}, 
Kodama has pointed out that the
exponential of the Chern-Simons action, 
which is equivalent to the 
Einstein action\cite{3-d gravity}, 
is an exact solution of the Hamiltonian 
constraint in the holomorphic representation of 
the Ashtekar formalism\cite{Ashtekar}. 
Connections with our solution are unclear.}
\begin{equation}
S_1[h] = 
\int d^3x \sqrt{h(x)}~ \{ a_1 +a_2 R(x) \} \ ,
\label{502}
\end{equation}
where
\begin{equation}
\begin{array}{lcl}
a_1 &=& \displaystyle{ \frac{16 (4 \pi)^{3/2}}{7 \mu^3 } 
        \biggl\{ \Lambda 
       + \frac{9 \mu^2 }{5 (11-36~ \xi) } 
       \biggr\}  } \  , \nonumber\\
& & \nonumber\\
a_2 &=& \displaystyle{
\frac{72 (4 \pi)^{3/2}}{(11-36~ \xi)\mu^3 } }\ .
\end{array}
\label{503}
\end{equation}
This result agrees with 
the one obtained in Ref.\cite{analysis} with 
$\xi = 0$. One may try to find other solutions to $S_1[h]$. 
It turns out 
that other solutions to $S_1[h]$, if exist, 
may be given by the integral of 
nonlocal functions of $h_{ij}$, i.e., $S_1[h]$ 
includes infinitely many higher
derivative terms. 

We would like to briefly discuss the renormalizability. 
In our formulation, 
the renormalizability of the theory requires 
that all physical quantities 
must be independent of the arbitrary mass parameter $\mu$, 
so that the Newton 
\lq\lq constant" $G$ and the cosmological 
\lq\lq constant" $\Lambda$ should be regarded 
as functions of $\mu$. To the first order, 
the above statement may be 
replaced by saying that the wave functional 
is independent of $\mu$. This is achieved by requiring that
the following combinations are independent of $\mu$:
\begin{equation}
\left.
\begin{array}{l}
G^2(\mu)~ \mu^3 \vspace{2mm} \\ 
\Lambda(\mu) +
\displaystyle{\frac{9~ \mu^2 }{5 (11-36~ \xi)}} 
\end{array}
\right\} \mu-{\rm independent} ~~~. 
\label{504}
\end{equation}
Thus the $\mu$-dependence can be absorbed into the redefinition 
of $G$ and $\Lambda$ to the first order in the strong coupling 
expansion. The above observation (\ref{504}) is a good news for
our strong coupling expansion, 
which is expected to give a good approximation scheme for 
large $\mu$ as discussed in the previous section, because 
the actual dimensionless expansion parameter 
$( G^2(\mu) \mu^4)^{-1}$ tends to zero as the mass scale $\mu$ 
increases. 

We will finally look for a solution to $S_1[h]$ 
in the case of the second solution (\ref{409}). 
Then, the first order equation (\ref{406}) is 
reduced to 
\begin{equation}
\Delta_{R}(x)\, S_1[h] 
  + \frac{7 \mu^3}{6 (4 \pi)^{3/2}} 
    \frac{\delta S_1[h]}{\delta {h_i}^i(x)} 
  = -  \sqrt{h(x)}  \Bigl( R(x) + 2 \Lambda \Bigr) \ .
\label{509}
\end{equation}
It is easy to show that a solution to (\ref{509}) is given by 
\begin{equation}
S_1[h] = 
\int d^3x \sqrt{h(x)}~ \{ {a'}_1 +{a'}_2 R(x) \} \,
\label{510}
\end{equation}
where
\begin{equation}
\begin{array}{lcl}
{a'}_1 &=& \displaystyle{- \frac{16 (4 \pi)^{3/2}}{7 \mu^3}
      \biggl[ \Lambda 
     - \frac{9 \mu^2}
     {5 (31 + 36~ \xi)} \biggr] } \ , \nonumber\\
 & & \nonumber\\
{a'}_2 &=& \displaystyle{ 
- \frac{72 (4 \pi)^{3/2}}{(31 + 36~ \xi) \mu^3}} \ .
\end{array}
\label{511}
\end{equation}
The difference between the solutions (\ref{502}) 
and (\ref{510}) lies only 
in the numerical coefficients and hence all 
the discussions on the solution 
(\ref{502}) will remain true for the solution (\ref{510}). 
 
%%%%%%%%%%%%%%%%%%%%%%%%%%%%%%%%%%%%%%%%%%%%%%%%%%%%%%%%%%%%%
\section{THE SECOND ORDER SOLUTIONS}

In the previous section, we have found the first order solution 
for $S_1[h]$ in the strong coupling expansion. For our expansion 
scheme to be consistent, we need to show that $S_n[h]$ for 
$n \geq 2 $ can in principle be constructed order by order 
and also that the successive terms obey, in particular, 
the relation (\ref{412}); otherwise our solutions are
meaningless. To this end, in this section we look for solutions 
to the second order equation 
\begin{equation}
\Delta_{R}(x)\, S_2[h] 
  -  G_{ijkl}(x) \biggl\{ 
  2 \frac{\delta S_0[h]}{\delta h_{kl}(x)} 
    \frac{\delta S_2[h]}{\delta h_{ij}(x)} 
  + \frac{\delta S_1[h]}{\delta h_{kl}(x)} 
    \frac{\delta S_1[h]}{\delta h_{ij}(x)}   \biggr\}= 0\  ,
\label{601}
\end{equation}
and show that $S_2[h]$ has the form (\ref{414}), as announced in 
Sec.IV. We also discuss 
a problem of finding higher order solutions.
The equation (\ref{601}) may be solved 
by assuming $S_2[h]$ to take the form 
\begin{equation}
S_2[h] = \int d^3x \sqrt{h(x)} \{ b_1 + b_2 R(x) 
          + b_3 R^2(x) + b_4 R_{ij}(x) R^{ij}(x) \} \ .  
\label{602}
\end{equation}
Note that the Riemann curvature $R_{ijkl}$ 
can be expressed in terms of 
$R$ and $R_{ij}$  in three-dimensions and also that the term  
$\int d^3x \sqrt{h} \nabla\!_i\, \nabla\!^i\, R$ 
is not included in $S_2[h]$ 
because it vanishes identically (without boundaries). 
Substitution of 
(\ref{408}), (\ref{502}) and (\ref{602}) into (\ref{601}) 
leads to
\begin{equation}
B_1 + B_2 R(x) + B_3 R^2(x) 
 + B_4 R_{ij}(x) R^{ij}(x) 
 + B_5 \nabla\!_i\, \nabla\!^i\, R(x) = 0 \ ,
\label{603}
\end{equation}
where
\begin{equation}  
\begin{array}{lcl}
B_1 &=&\displaystyle{
        b_1 \alpha_1   \frac{\mu^3}{3  (4 \pi)^{3/2}} 
      + b_2 \beta_1    \frac{\mu^5}{30 (4 \pi)^{3/2}} 
      + ( b_3 \gamma_1 + b_4 {\gamma'}_1 )
                       \frac{\mu^7}{840 (4 \pi)^{3/2}}
      + \frac{3}{8} (a_1)^2  } \ , \nonumber \vspace{2mm}\\
B_2 &=&\displaystyle{
        b_2 \beta_2    \frac{\mu^3}{3  (4 \pi)^{3/2}} 
      + ( b_3 \gamma_2 + b_4 {\gamma'}_2 ) 
                       \frac{\mu^5}{30 (4 \pi)^{3/2}}
      + \frac{1}{4}  a_1 a_2  } \ , \nonumber\vspace{2mm}\\
B_3 &=& \displaystyle{
       ( b_3 \gamma_3 + b_4 {\gamma'}_3 )
                       \frac{\mu^3}{3 (4 \pi)^{3/2}}
      + \frac{3}{8} (a_2)^2   }\ , \nonumber\vspace{2mm}\\
B_4 &=& \displaystyle{
       ( b_3 \gamma_4 + b_4 {\gamma'}_4 ) 
                       \frac{\mu^3}{3 (4 \pi)^{3/2}}
      - (a_2)^2  } \ , \nonumber\vspace{2mm}\\
B_5 &=& \displaystyle{
       ( b_3 \gamma_5 + b_4 {\gamma'}_5 )
                       \frac{\mu^3}{3 (4 \pi)^{3/2}}} 
      \ .
\end{array}
\label{604}
\end{equation}
The equation (\ref{603}) or equivalently the five equations 
\begin{equation}
B_n = 0  \ ,~~~~~~ {\rm for} ~~~ n = 1,2,\cdots,5 \ , 
\label{605}
\end{equation}
will, in general, have no solutions 
because the number of the free parameters 
($b_m, m= 1,\cdots,4$) is less 
than the number of the equations 
($B_n, n= 1,\cdots,5$). This is due to the fact that the term 
$\nabla\!_i\, \nabla\!^i\, R(x) $ 
appears in (\ref{603}) but not in (\ref{602}), 
in other words, the Wheeler-DeWitt equation 
is a local equation imposed by 
the Hamiltonian density (not the Hamiltonian itself), 
while $S_2[h]$ is 
assumed to be the three-dimensional integral 
of local functions\footnote{ 
This kind of problems does not occur 
for Yang-Mills theory because the 
Hamiltonian appears in the Schr\"odinger equation.}. 
Hence we need one more parameter 
to solve the equation (\ref{603}) or (\ref{605}). 
To this end, we will here use 
the arbitrariness of defining the kernel 
$K_{i'j'kl}$. It is not difficult to show 
that the equation (\ref{603}) 
or (\ref{605}) has a solution, provided 
that $\xi$ is chosen to be one of 
the solutions to the equation 
\begin{equation}
9600~ \xi^2 - 7633~ \xi + 196 = 0 \ . 
\label{606}
\end{equation}

To solve higher order equations for 
$S_n[h] (n \geq 3) $ with the ansatz of locality, 
we will face a similar problem and need 
to generalize the heat equation (\ref{205}) 
to include more arbitrary parameters. Thus, 
in the strong coupling expansion the
form of the wave functional to the second or 
higher order crucially depends on the 
choice of the heat equation. We do not know whether 
or not this fact causes 
serious problems in our formulation because 
the wave functional itself is not
a physical observable and because it is inevitable 
that the wave functional depends on
the renormalization prescription. Our hope is 
that physical quantities can be 
independent of our renormalization prescription 
and that our solutions have 
important physical meanings with a special 
choice of the heat equation. 

In the case of the second solution (\ref{409}) 
to the zeroth order, we can also 
show that the second order equation (\ref{601}) 
with the ansatz (\ref{602}) has 
a solution, provided that $\xi$ is chosen to be 
one of the solutions to the 
equation 
\begin{equation}
40800~ \xi^2 - 9307~ \xi + 1834 = 0 \ . 
\label{608}
\end{equation}
Since the discriminant is negative, the solutions are complex. 
This implies that 
an imaginary part appears in the renormalized 
Hamiltonian operator through the 
kernel $K_{i'j'kl}$, so the second solution (\ref{409}) 
to the zeroth order 
seems to lead to an undesirable result. 
Since the hermiticity of the Hamiltonian 
is unclear in quantum gravity due to the 
lack of the knowledge of the functional 
measure ${\cal D}h$, we will not discuss 
the problem furthermore.

%%%%%%%%%%%%%%%%%%%%%%%%%%%%%%%%%%%%%%%%%%%%%%%%%%%%%%%%%%%
\section{CONCLUSION}

We have proposed a renormalization prescription 
to the Wheeler-DeWitt equation
and solved it to the second order 
in the strong coupling expansion to study quantum 
geometry at length scales much smaller 
than the Planck length. 
We have restricted our attention to a class of local 
solutions for mainly technical reasons. 
Our formulation is not, however, limited to a class
of local solutions and our renormalization prescription 
does not rely on the strong coupling expansion.
It would be challenging to look for non-local solutions 
in the strong coupling expansion and for exact ( local 
or non-local ) solutions without the expansion\cite{Kowalski}.

We have also restricted our attention 
to pure gravity. Recently, the dilaton gravity
$$
{\cal L} = \sqrt{g}~ {\rm e}^{- 2 \Phi} 
           \Bigl( R - 4 D_{\mu} \Phi D^{\mu} \Phi 
           + \frac{1}{12} H_{\mu\nu\rho} H^{\mu\nu\rho} \Bigr) 
$$
has extensively been studied to reveal 
stringy phenomena\cite{dilaton}. 
It would be of great interest to investigate 
the theory in our formulation 
to reveal stringy dynamics beyond the Planck scale. 

\ \ 

\noindent
{\large\bf ACKNOWLEDGMENTS} 

We would like to thank A. Hosoya, K. Inoue, 
M. Kato, Y. Kazama, C.S. Lim,
J. Soda and T. Yoneya for useful discussions. 
We also thank T. Horiguchi for
collaboration at an early stage of this work.

\newpage

%%%%%%%%%%%%%%%%%%%%%%%%%%%%%%%%%%%%%%%%%%%%%%%%%%%%%%%%%%%%%%%
\appendix
\renewcommand{\thesection}{%
   \bf{APPENDIX~} \Alph{section}:}
\renewcommand{\theequation}{% 
    \Alph{section}.\arabic{equation}}

\section{{\normalsize\bf THE HEAT KERNEL IN THE COINCIDENCE LIMIT}}

In this appendix, we will 
compute the heat kernel $K_{i'j'kl}$ and 
its covariant derivatives $K_{i'j'kl;a'},K_{i'j'kl;a'b'}$ and 
$K_{i'j'kl;a'b'c'd'}$ 
in the coincidence limit $x' \rightarrow x$. 

According to the standard technique\cite{Schwinger-DeWitt},
the heat equation (\ref{205}) 
with the initial condition (\ref{206}) 
can be solved by the ansatz 
\begin{equation}
K_{i'j'kl}(x',x;t) = 
    \frac{ (\Delta (x',x))^{1/2} }{ (4 \pi t)^{3/2} } ~
    {\rm e}^{ - \frac{ \sigma(x',x) }{2 t} } ~
    \sum^{\infty}_{n=0} a^{(n)}_{i'j'kl}(x',x)~t^n \ ,
\label{a3}
\end{equation}
where the bi-scalar $\sigma(x',x)$ 
is the geodesic integral equal to 
one half the square of the distance along 
the geodesic between $x'$ 
and $x$, and the bi-scalar $\Delta(x',x)$ 
is defined by\footnote{ The 
$\sigma_{;i'j}(x',x)$ denotes 
the covariant derivatives of $\sigma(x',x)$ 
with respect to ${x'}^i$ and $x^j$. 
The bi-scalar $\Delta(x',x)$ should 
not be confused with the differential operator 
$\Delta(x)$ or $\Delta(x;t)$ 
in the text.}
\begin{equation}
\Delta(x',x) = 
 h(x')^{- \frac{1}{2}}~ 
 {\rm det}(\sigma_{;i'j}(x',x))~ h(x)^{- \frac{1}{2}} \ .
\label{a4}
\end{equation}
The properties of $\sigma(x',x)$ and $\Delta(x',x)$ 
have been discussed in 
detail in Ref.\cite{dynamical theory}. We will not
 repeat the discussions 
here. 
 
The initial condition (\ref{206}) implies that 
\begin{equation}
\lim_{x' \rightarrow x} a^{(0)}_{i'j'kl} = \sqrt{h} G_{i'j'kl} \ . 
\label{a5}
\end{equation}
Inserting the ansatz (\ref{a3}) into 
the equation (\ref{205}), one finds 
\begin{equation}
\sigma^{;p'}(x',x) a^{(0)}_{i'j'kl;p'}(x',x) = 0 \ ,
\label{a6}
\end{equation}
\begin{equation}
\begin{array}{l}
n~ a^{(n)}_{i'j'kl}(x',x) 
+ \sigma^{;p'}(x',x) ~ a^{(n)}_{i'j'kl;p'}(x',x)  \\
\\
= \Bigl(\Delta(x',x) \Bigr)^{-\frac{1}{2}} 
    ~ { \Bigl( (\Delta(x',x))^{\frac{1}{2}} 
    a^{(n-1)}_{i'j'kl}(x',x) \Bigr)^{;p'}}_{p'}
 + \xi R(x') a^{(n-1)}_{i'j'kl}(x',x) \ , \vspace{1mm} \\
\hspace{7cm} {\rm for}\  ~~~~ n =1,2,\cdots \ .
\end{array}
\label{a7}
\end{equation}
For our purpose, we need to know values 
for the first few $a$'s in the 
expansion (\ref{a3}) and some of 
their covariant derivatives in the 
limit $x' \rightarrow x$. 
They can be obtained by repeatedly differentiating 
the equations (\ref{a6}) and (\ref{a7}) 
and then by taking the limit 
$x' \rightarrow x$. 
The results are 
\begin{eqnarray}
\lim_{x' \rightarrow x}  a^{(0)}_{i'j'kl}~~~~~~~~      
   &=& \lefteqn{ \sqrt{h} G_{i'j'kl} ~~~  ,}     \nonumber \\
\lim_{x' \rightarrow x} a^{(0)}_{i'j'kl;a'}~~~~~~  
   &=&  0  ~~~  ,                 \nonumber \\
\lim_{x' \rightarrow x} a^{(0)}_{i'j'kl;a'b'}~~~~  
   &=&  \sqrt{h} G_{p'j'kl} 
        \Bigl( - \frac{1}{2}{R^{p'}}_{i'a'b'} \Bigr)
      + ( i' \leftrightarrow j' )~~~   ,   \nonumber \\
\lim_{x' \rightarrow x} a^{(0)}_{i'j'kl;a'b'c'}~~ 
    &=&  \sqrt{h} G_{p'j'kl} 
         \Bigl( - \frac{1}{3} {R^{p'}}_{i'a'b';c'}  
           - \frac{1}{3} {R^{p'}}_{i'a'c';b'} \Bigr) 
      + ( i' \leftrightarrow j' )~~~   ,       \nonumber \\
%%%%%%%%%%%%%%%%%%%%%%%%%%%%%%%%%%%%%%%%%%%%%%%%%%%%%%%%%%%%%%%%%
\lim_{x' \rightarrow x}a^{(0)}_{i'j'kl;a'b'c'd'}   
   &=&  \sqrt{h} G_{p'q'kl} \Bigl( \frac{1}{4} 
    ( {R^{p'}}_{i'a'b'} {R^{q'}}_{j'c'd'} 
    + {R^{p'}}_{i'a'c'} {R^{q'}}_{j'b'd'}
    + {R^{p'}}_{i'a'd'} {R^{q'}}_{j'b'c'}  \nonumber\\
&&\hspace{20mm}
    + {R^{p'}}_{i'c'd'} {R^{q'}}_{j'a'b'}
    + {R^{p'}}_{i'b'd'} {R^{q'}}_{j'a'c'}
    + {R^{p'}}_{i'b'c'} {R^{q'}}_{j'a'd'} )  \Bigr) \nonumber\\ 
&&  + \Bigl\{   \sqrt{h} G_{pj'kl} \Bigl( \frac{1}{12} (
      {R^{p'}}_{i'q'a'} {R^{q'}}_{b'd'c'} 
      + {R^{p'}}_{i'q'a'} {R^{q'}}_{d'b'c'} 
\nonumber\\ 
&&\hspace{26mm}
    + {R^{p'}}_{i'q'b'} {R^{q'}}_{a'c'd'} 
    + {R^{p'}}_{i'q'b'} {R^{q'}}_{c'a'd'} 
\nonumber\\   
&&\hspace{26mm}
    + {R^{p'}}_{i'q'c'} {R^{q'}}_{a'b'd'} 
    + {R^{p'}}_{i'q'c'} {R^{q'}}_{b'a'd'} 
\nonumber\\  
&&\hspace{26mm}
    + {R^{p'}}_{i'q'd'} {R^{q'}}_{a'b'c'} 
    + {R^{p'}}_{i'q'd'} {R^{q'}}_{b'a'c'}  ) 
\nonumber\\ 
&&\hspace{22mm} 
    +  \frac{1}{8} (       
      {R^{q'}}_{i'a'b'} {R^{p'}}_{q'c'd'} 
    + {R^{q'}}_{i'a'c'} {R^{p'}}_{q'b'd'}
    + {R^{q'}}_{i'a'd'} {R^{p'}}_{q'b'c'}     \nonumber\\ 
&&\hspace{26mm}
    + {R^{q'}}_{i'c'd'} {R^{p'}}_{q'a'b'}
    + {R^{q'}}_{i'b'd'} {R^{p'}}_{q'a'c'}
    + {R^{q'}}_{i'b'c'} {R^{p'}}_{q'a'd'}   )  \nonumber\\ 
&&\hspace{22mm}
    -  \frac{1}{4}~ ({R^p}_{i'a'b';c'd'} 
    + {R^p}_{i'a'c';b'd'} + {R^p}_{i'a'd';b'c'} )  
    + ( i'\leftrightarrow j')   \Bigr\}     \ ,     \nonumber\\ 
%%%%%%%%%%%%%%%%%%%%%%%%%%%%%%%%%%%%%%%%%%%%%%%%%%%%%%%%%%%%%%%%%%
\lim_{x' \rightarrow x} a^{(1)}_{i'j'kl}~~~~~~~~        
   &=& \sqrt{h} G_{i'j'kl} 
   \Bigl( -\frac{1}{6} + \xi \Bigr) R  \ ,  \nonumber\\ 
%%%%%%%%%%%%%%%%%%%%%%%%%%%%%%%%%%%%%%%%%%%%%%%%%%%%%%%%%%%%%%%%%%
\lim_{x' \rightarrow x} a^{(1)}_{i'j'kl;a'}~~~~~~ 
   &=& \sqrt{h}  G_{i'j'kl} 
       \Bigl( - \frac{1}{12} + \frac{1}{2}~ \xi \Bigr) R_{;a'}    
     + \Bigl\{ \sqrt{h} G_{q'j'kl} 
     \Bigl( \frac{1}{6} {{R^{q'}}_{i'a'p'}}^{;p'} \Bigr)
              + ( i'\leftrightarrow j') \Bigr\} \ ,  \nonumber\\ 
%%%%%%%%%%%%%%%%%%%%%%%%%%%%%%%%%%%%%%%%%%%%%%%%%%%%%%%%%%%%%%%%%%%
\lim_{x' \rightarrow x} a^{(1)}_{i'j'kl;a'b'}~~~~  
   &=& \sqrt{h} G_{i'j'kl} 
       \Bigl(   \frac{1}{90} R^{p'q'} R_{p'a'q'b'} 
              - \frac{1}{45} {R^{p'}}_{a'} R_{p'b'}
              + \frac{1}{90} {R^{p'q'm'}}_{a'} R_{p'q'm'b'}   
               \nonumber\\ 
&&\hspace{18mm}
              - \Bigl(\frac{1}{20} 
              - \frac{1}{3}~ \xi \Bigr) R_{;a'b'} 
              -  \frac{1}{60} {R_{a'b';p'}}^{p'}   \Bigr)    
                \nonumber\\ 
&& + \sqrt{h} G_{p'q'kl} 
        \Bigl( \frac{1}{6} {R^{p'}}_{i'm'a'} 
               {{{R^{q'}}_{j'}}^{m'}}_{b'}  
             + \frac{1}{6} {R^{p'}}_{j'm'a'}
              {{{R^{q'}}_{i'}}^{m'}}_{b'} \Bigr)   
      \nonumber\\ 
&& + \Bigl\{
     \sqrt{h} G_{p'j'kl} 
        \Bigl(  \Bigl( \frac{1}{12} 
     - \frac{1}{2}~ \xi \Bigr) R {R^{p'}}_{i'a'b'}
     + \frac{1}{12}  R_{q'i'm'a'} {R^{p'q'm'}}_{b'}
               \nonumber\\ 
&&\hspace{25mm}   
     + \frac{1}{12}  R_{q'i'm'b'} {R^{p'q'm'}}_{a'}        
     -  \frac{1}{12} {{{R^{p'}}_{i'}}^{q'}}_{a';q'b'}  
     -  \frac{1}{12} {{{R^{p'}}_{i'}}^{q'}}_{b';q'a'} \Bigr) 
      \nonumber\\ 
&&\hspace{5mm}    
     + ( i'\leftrightarrow j')   \Bigr\}   \ ,                  
     \nonumber\\ 
%%%%%%%%%%%%%%%%%%%%%%%%%%%%%%%%%%%%%%%%%%%%%%%%%%%%%%%%%%%%%%%%
\lim_{x' \rightarrow x} a^{(2)}_{i'j'kl} ~~~~~~~~  
   &=& \sqrt{h} G_{i'j'kl} \Bigl(   
     \Bigl( \frac{1}{72}- \frac{1}{6} ~\xi 
    + \frac{1}{2}~ \xi^2 \Bigr) R^2 
   - \frac{1}{180} R_{p'q'} R^{p'q'}  \nonumber\\ 
&&\hspace{18mm}
    + \frac{1}{180} R^{p'q'm'n'} R_{p'q'm'n'}  
    - \Bigl( \frac{1}{30} - \frac{1}{6}~ \xi \Bigr) {R_{;p'}}^{p'}
      \Bigr)     \nonumber\\ 
&& + \frac{1}{6} \sqrt{h} G_{p'q'kl} 
       {R^{p'}}_{i'm'n'} {{R^{q'}}_{j'}}^{m'n'}  \nonumber\\ 
&& + \Bigl\{ \frac{1}{12} 
             \sqrt{h} G_{p'j'kl} R_{q'i'm'n'} R^{p'q'm'n'}  
   + ( i'\leftrightarrow j') \Bigr\} \ .  \label{a8} 
\end{eqnarray} 

These results enable us to compute $K_{i'j'kl}$, $K_{i'j'kl;a'}$, 
$K_{i'j'kl;a'b'}$ and $K_{i'j'kl;a'b'c'd'}$ 
in the limit $x' \rightarrow x$.
\begin{eqnarray} 
\lefteqn{ \lim_{x' \rightarrow x}K_{i'j'kl}    
   =   \frac{1}{(4\pi t)^{3/2}} \sqrt{h} G_{i'j'kl} 
       \Bigl\{~ 1 - t \Bigl( \frac{1}{6} - \xi \Bigr) R ~ \Bigr\} 
       + O(t^{\frac{1}{2}})  \ , } &&  
\label{a9} \\
\lefteqn{\lim_{x' \rightarrow x} K_{i'j'kl;a'}   } &&
\nonumber\\
  && =   \frac{1}{(4\pi)^{3/2} t^{1/2}} 
         \Bigl\{~  
         - \frac{1}{2} \Bigl(\frac{1}{6} - \xi \Bigr) 
           \sqrt{h} G_{i'j'kl} R_{;a'} 
\nonumber\\
   &&\hspace{24mm}  
  + \frac{1}{6} 
    \Bigl(~ \sqrt{h} G_{p'j'kl} {{R^{p'}}_{i'a'q'}}^{;q'}
  + ( i' \leftrightarrow j' ) ~ \Bigr) ~\Bigr\}
        + O(t^{\frac{1}{2}})  \ ,  
\label{a10}  \\
\lefteqn{\lim_{x' \rightarrow x} K_{i'j'kl;a'b'}  } & &\nonumber \\
  && =     \frac{1}{(4\pi)^{3/2} t^{5/2}} 
      \Bigl\{~ - \frac{1}{2} 
                 \sqrt{h} G_{i'j'kl} h_{a'b'} \Bigr\} \nonumber\\
  &&~ + ~ \frac{1}{(4 \pi t )^{3/2}} ~ 
        \Bigl\{~ \sqrt{h} G_{ijkl} \Bigl(- \frac{1}{6}  R_{a'b'} 
         + \Bigl(\frac{1}{12} 
               - \frac{1}{2}~ \xi \Bigr) h_{a'b'} R \Bigr)
   \nonumber \\ 
   & &\hspace{20mm} ~~~
       - \frac{1}{2} \Bigl(~ \sqrt{h} G_{p'j'kl} {R^{p'}}_{i'a'b'} 
            + ( i' \leftrightarrow j' ) ~\Bigr)   ~ \Bigr\} 
             + O(t^{-\frac{1}{2}})   \ , 
\label{a11} \\
\lefteqn{\lim_{x' \rightarrow x} K_{i'j'kl;a'b'c'd'} } 
\nonumber\\  &&  
    =   \frac{1}{ (4\pi t)^{3/2} } \Bigl\{ ~
         \frac{1}{t^2} F^{(0)}_{i'j'kla'b'c'd'}
       + \frac{1}{t}   F^{(1)}_{i'j'kla'b'c'd'}  
       +               F^{(2)}_{i'j'kla'b'c'd'} ~ \Bigr\}    
     + O(t^{-\frac{1}{2}})  \ ,   \label{a12}   
\end{eqnarray}
where
\begin{eqnarray} 
\lefteqn{ F^{(0)}_{i'j'kla'b'c'd'}  
   =  \frac{1}{8}\sqrt{h} G_{i'j'kl} 
     [ h_{\underline{a'}\underline{b'}} 
       h_{\underline{c'}\underline{d'}}]_s  \ ,} && \nonumber\\
\lefteqn{ F^{(1)}_{i'j'kla'b'c'd'} } && \nonumber\\
&& = \sqrt{h} G_{i'j'kl} \Bigl\{ 
         - \frac{1}{6} ( R_{a'c'b'd'} + R_{a'd'b'c'} )  
         + \frac{1}{12} 
             [ h_{\underline{a'}\underline{b'}} 
               R_{\underline{c'}\underline{d'}} ]_s
         - \Bigl(\frac{1}{48} -  \frac{1}{8}~ \xi \Bigr) 
              R~ [ h_{\underline{a'}\underline{b'}} 
                   h_{\underline{c'}\underline{d'}}]_s  ~\Bigr\} 
\nonumber\\ 
&&\hspace{5mm}
  + \Bigl\{~ \frac{1}{4} \sqrt{h} G_{p'j'kl} 
           [ h_{\underline{a'}\underline{b'}}
            {R^{p'}}_{i'\underline{c'}\underline{d'}} ]_s 
       + ( i' \rightarrow j') ~ \Bigr\} \ , \nonumber\\
\lefteqn{F^{(2)}_{i'j'kla'b'c'd'} } && \nonumber \\
&& = \sqrt{h} G_{i'j'kl} \biggl\{ 
     \frac{1}{180} 
             \Bigl( R_{p'a'} ( 11 {R^{p'}}_{c'b'd'} 
                             - 10 {R^{p'}}_{d'b'c'} )
                  + R_{p'b'} ( 11 {R^{p'}}_{c'a'd'} 
                             - 10 {R^{p'}}_{d'a'c'} )   
\nonumber\\
&& \hspace{30mm}  + R_{p'c'} ( 11 {R^{p'}}_{b'a'd'}
                             - 10 {R^{p'}}_{d'a'b'} ) 
                  + R_{p'd'} ( 11 {R^{p'}}_{b'a'c'}
                             - 10 {R^{p'}}_{c'a'b'} )  \Bigr) 
\nonumber\\                   
&& \hspace{23mm}   
   + \Bigl( \frac{1}{36} - \frac{1}{6}~ \xi\Bigr)
       R ~( R_{a'c'b'd'} + R_{a'd'b'c'} )  
   + \frac{1}{72}  [ R_{\underline{a'}\underline{b'}} 
                     R_{\underline{c'}\underline{d'}}]_s 
 \nonumber\\
&&\hspace{23mm}
    + \frac{1}{8} \biggl(  
        \Bigl(  \frac{1}{72} - \frac{1}{6}~\xi 
              + \frac{1}{2} ~\xi^2  \Bigr) R^2 
        - \frac{1}{180} R^{p'q'} R_{p'q'} 
       + \frac{1}{180} R^{p'q'm'n'} R_{p'q'm'n} 
  \nonumber\\
&&\hspace{30mm}
     - ( \frac{1}{30} - \frac{1}{6} \xi ) {R^{;p'}}_{p'} \biggr)
             [ h_{\underline{a'}\underline{b'}} 
               h_{\underline{c'}\underline{d'}}]_s 
   - \Bigl( \frac{1}{72} - \frac{1}{12}~ \xi \Bigr) 
          R~ [h_{\underline{a'}\underline{b'}} 
              R_{\underline{c'}\underline{d'}}]_s
\nonumber\\ 
&&\hspace{23mm}
 - \frac{1}{72}[ {R^{p'q'}}_{\underline{a'}\underline{b'}} 
                  R_{p'q'\underline{c'}\underline{d'}} ]_s 
 + \frac{1}{45}[ {R^{p'q'}}_{\underline{a'}\underline{b'}} 
                  R_{p'\underline{c'} q'\underline{d'}} ]_s  
 + \frac{1}{90}[ {{{R^{p'}}_{\underline{a'}}}^{q'}}_{\underline{b'}} 
                  R_{p'\underline{c'} q'\underline{d'}} ]_s  
\nonumber\\                   
&& \hspace{23mm}
-   \frac{1}{180} 
        [ h_{\underline{a'}\underline{b'}} 
          R^{p'q'} R_{p'\underline{c'} q'\underline{d'} }  ]_s 
+   \frac{1}{90} 
        [ h_{\underline{a'}\underline{b'}}  
         {R^{p'}}_{\underline{c'}} R_{p'\underline{d'}} ]_s 
- \frac{1}{180} 
        [ h_{\underline{a'}\underline{b'}}  
         {R^{p'q'm'}}_{\underline{c'}} R_{p'q'm'\underline{d'}} ]_s  
\nonumber\\                   
&& \hspace{23mm}             
- \frac{1}{20}  
        [ R_{\underline{a'}\underline{b'};
             \underline{c'}\underline{d'}}]_s 
+ \frac{1}{120} 
        [ h_{\underline{a'}\underline{b'}} 
         {R_{\underline{c'}\underline{d'};p'}}^{p'}]_s 
+ \Bigl(\frac{1}{40} -\frac{1}{6} ~\xi \Bigr)  
        [ h_{\underline{a'}\underline{b'}} 
          R_{;\underline{c'}\underline{d'}}]_s \biggr\}
\nonumber\\                   
&& + \sqrt{h} G_{p'q'kl} \biggl\{ 
 - \frac{1}{12} 
 [ h_{\underline{a'}\underline{b'}} {R^{p'}}_{i'm'\underline{c'}} 
   {{{R^{q'}}_{j'}}^{m'}}_{\underline{d'}} ]_s   
 - \frac{1}{12} 
 [ h_{\underline{a'}\underline{b'}} {R^{p'}}_{i'm'\underline{d'}} 
  {{{R^{q'}}_{j'}}^{m'}}_{\underline{c'}} ]_s   
\nonumber\\                   
&& \hspace{23mm} 
    + \frac{1}{48}{{R^{p'}}_{i'}}^{m'n'} 
       {R^{q'}}_{j'm'n'}[ h_{\underline{a'}\underline{b'}} 
                          h_{\underline{c'}\underline{d'}}]_s   
    + \frac{1}{4} 
     [ {R^{p'}}_{i'\underline{a'}\underline{b'}} 
       {R^{q'}}_{j'\underline{c'}\underline{d'}} ]_s    \biggr\} 
\nonumber\\                   
&& + \biggl\{ \sqrt{h} G_{pj'kl} \Bigl\{ ~ 
\frac{1}{12} (
      {R^{p'}}_{i'q'a'} {R^{q'}}_{b'd'c'} 
    + {R^{p'}}_{i'q'a'} {R^{q'}}_{d'b'c'}    
    + {R^{p'}}_{i'q'b'} {R^{q'}}_{a'c'd'} 
    + {R^{p'}}_{i'q'b'} {R^{q'}}_{c'a'd'}  
     \nonumber\\  
&&\hspace{29mm}
    + {R^{p'}}_{i'q'c'} {R^{q'}}_{a'b'd'} 
    + {R^{p'}}_{i'q'c'} {R^{q'}}_{b'a'd'}  
    + {R^{p'}}_{i'q'd'} {R^{q'}}_{a'b'c'} 
    + {R^{p'}}_{i'q'd'} {R^{q'}}_{b'a'c'}  ) 
     \nonumber\\
&& \hspace{23mm} -  \frac{1}{4}   ({R^{p'}}_{i'a'b';c'd'} 
                      + {R^{p'}}_{i'a'c';b'd'} 
                      + {R^{p'}}_{i'a'd';b'c'} )     \nonumber\\
&& \hspace{23mm}
 + \frac{1}{96}  
 R^{p'q'm'n'} R_{q'i'm'n'} ~
         [ h_{\underline{a'}\underline{b'}} 
           h_{\underline{c'}\underline{d'}} ]_s 
 - \Bigl(\frac{1}{24} - \frac{1}{4} ~\xi \Bigr) 
 R~ [h_{\underline{a'}\underline{b'}}
    {R^{p'}}_{i'\underline{c'}\underline{d'}} ]_s 
\nonumber\\ 
&& \hspace{23mm}
 - \frac{1}{24}[ h_{\underline{a'}\underline{b'}} 
                {R^{p'q'm'}}_{\underline{d'}} 
                 R_{q'i'm'\underline{c'}} ]_s  
 - \frac{1}{24}[ h_{\underline{a'}\underline{b'}} 
                {R^{p'q'm'}}_{\underline{c'}} 
                 R_{q'i'm'\underline{d'}} ]_s  
\nonumber\\ 
&& \hspace{23mm}
  + \frac{1}{12}  
     [ R_{\underline{a'}\underline{b'}}
       {R^{p'}}_{i'\underline{c'}\underline{d'}}]_s
  + \frac{1}{8} 
     [ {R^{q'}}_{i'\underline{a'}\underline{b'}} 
       {R^{p'}}_{q'\underline{c'}\underline{d'}}]_s  
  + \frac{1}{24}
     [ h_{\underline{a'}\underline{b'}} 
     {{{R^{p'}}_{i'}}^{q'}}_{\underline{c'};q\underline{d'}} ]_s  
\nonumber\\ 
&& \hspace{23mm}
  + \frac{1}{24}
     [ h_{\underline{a'}\underline{b'}} 
      {{{R^{p'}}_{i'}}^{q'}}_{\underline{d'};q\underline{c'}} ]_s  
     \Bigr\}
  ~+~ ( i' \leftrightarrow j')   \biggr\}   \ , \\
\label{a13}
\end{eqnarray}
Here, we have used a notation  
$[X_{\underline{a}\underline{b} ij\cdots} 
  Y_{\underline{c}\underline{d} kl\cdots}]_s$, 
which is an abbreviation of the sum of the six 
terms in different order with 
respect to the underlined indices  
$\underline{a}, \underline{b}, \underline{c}$ and 
$\underline{d}$, i.e.,
\begin{equation}
\begin{array}{lcl}
[ X_{\underline{a}\underline{b} ij\cdots} 
  Y_{\underline{c}\underline{d} kl\cdots} ]_s  & \equiv & 
   ~~ X_{abij\cdots}Y_{cdkl\cdots}
    + X_{cdij\cdots}Y_{abkl\cdots}  \\
&&  + X_{acij\cdots}Y_{bdkl\cdots}   
    + X_{bdij\cdots}Y_{ackl\cdots}  \\
&&  + X_{adij\cdots}Y_{bckl\cdots}   
    + X_{bcij\cdots}Y_{adkl\cdots}   \ .
\end{array}
\end{equation}
The indices which are not underlined are kept fixed 
in the original positions.

We have written the terms of order $t^{-1/2}$ 
in (\ref{a9}) and(\ref{a10}) 
because they play an important role 
in the computations of the anomalous 
term in (\ref{305}). As discussed in Sec. III, 
the anomaly free requirement 
leads to the condition (\ref{306}), 
and then all terms of order 
$t^{-1/2 +m} (m \geq 0)$ do not contribute 
to final results. The formulas 
(\ref{a9}) to (\ref{a13}) are enough 
to solve the Wheeler-DeWitt equation 
to the second order in the strong coupling expansion.

%%%%%%%%%%%%%%%%%%%%%%%%%%%%%%%%%%%%%%%%%%%%%%%%%%%%%%%%%%%%%%%%

\section{{\normalsize\bf THE FORMULAS (2.7) TO (2.10)}}

In this appendix, 
we will derive the equations (\ref{207}) to (\ref{210}) 
with the coefficients (\ref{211}) and (\ref{212}).

Substituting (\ref{204}) 
into the left hand side of (\ref{207}) and (\ref{208}), 
we find 
\begin{eqnarray}
\lefteqn{\Delta(x;t) \int d^3y \sqrt{h}  
= \lim_{x' \rightarrow x} K_{i'j'kl}(x',x;t) 
\textstyle{\frac{\sqrt{h(x)}}{4} } }\nonumber\\
&& \hspace{2cm} \times
\Bigl(  h^{i'j'}(x) h^{kl}(x) 
      - h^{i'k}(x) h^{j'l}(x) 
      - h^{i'l}(x) h^{j'k}(x) \Bigr) \ ,
\label{b1} \\
\lefteqn{\Delta(x;t) \int d^3y \sqrt{h} R } \nonumber\\
&&=\lim_{x' \rightarrow x} 
\Bigl( K_{i'j'kl}(x',x;t) I^{i'j'kl}(x) 
     + K_{i'j'kl;a'b'}(x',x;t) I^{i'j'kla'b'}(x) \Bigr) \ ,
\label{b2}
\end{eqnarray}
where
\begin{equation}
\begin{array}{lcl} 
I^{i'j'kl} &=& \displaystyle{\frac{\sqrt{h}}{2}}  \Bigl\{ 
    \displaystyle{
    \frac{1}{2}} h^{i'j'} h^{kl} R -  h^{i'k} h^{j'l} R  
   \vspace{3mm}\\
& &\hspace{5mm}  
     -  h^{i'j'} R^{kl} -  h^{kl} R^{i'j'} 
     + 2 h^{i'k} R^{j'l} + 2 h^{i'l} R^{j'k}  \Bigr\} \ , \\
I^{i'j'kla'b'} &=& \displaystyle{\frac{\sqrt{h}}{2}}   \Bigl\{
        h^{i'j'} h^{kl} h^{a'b'} - h^{i'j'} h^{kb'} h^{la'} 
      - h^{i'b'} h^{j'a'} h^{kl} \\
& &\hspace{5mm}
      - h^{i'k} h^{j'l} h^{a'b'} 
      + 2 h^{i'l} h^{j'b'} h^{ka'} \Bigr\} \ . 
\end{array}
\label{b3}
\end{equation}
Here, we have dropped surface terms. 
Using the results in Appendix A, we obtain 
\begin{eqnarray} 
\lefteqn{\Delta(x;t) \int d^3y \sqrt{h} 
 = \frac{\sqrt{h(x)}}{(4 \pi t)^{3/2}} 
 \Bigl\{ - \frac{21}{8} \Bigr\}
 + O(t^{-\frac{1}{2}})  \ , } && \hspace{12cm}
\label{b4} \\
\lefteqn{\Delta(x;t) \int d^3y \sqrt{h} R 
  = \frac{\sqrt{h(x)}}{(4 \pi t)^{3/2}}  \Bigl\{~ 
    \frac{1}{t} \Bigl( \frac{3}{2} \Bigr)
 + \Bigl( - \frac{11}{24} + \frac{3}{2}~ \xi \Bigr)  R(x)~ \Bigr\}
 + O(t^{-\frac{1}{2}})   \ . } && \hspace{12cm}
\label{b5} 
\end{eqnarray}

Computations of (\ref{209}) and (\ref{210}) 
are a little lengthy but straightforward. 
After some calculations, we have 
\begin{eqnarray}
\lefteqn{\Delta(x;t)  \int d^3y \sqrt{h} R^2  } \nonumber\\  
&& = \lim_{x' \rightarrow x} \Bigl\{ 
       K_{i'j'kl}(x',x;t)  J^{i'j'kl}(x)  
    +  K_{i'j'kl;a'}(x',x;t) J^{i'j'kla'}(x)    \nonumber\\   
 &&\hspace{11mm}
    +  K_{i'j'kl;a'b'}(x',x;t) J^{i'j'kla'b'}(x)  \nonumber\\
 &&\hspace{11mm}
    +  K_{i'j'kl;a'b'c'd'}(x',x;t)  
    J^{i'j'kla'b'c'd'}(x) \Bigr\} \ ,
\label{b6} \\
\lefteqn{\Delta(x;t)  \int d^3y \sqrt{h} R_{ij} R^{ij} }\nonumber\\  
&& = \lim_{x' \rightarrow x}  \Bigl\{ 
       K_{i'j'kl}(x',x;t) \overline{J}^{i'j'kl}(x)  
    +  K_{i'j'kl;a'}(x',x;t) 
    \overline{J}^{i'j'kla'}(x)  \nonumber\\      
&&\hspace{11mm}
    +  K_{i'j'kl;a'b'}(x',x;t) 
    \overline{J}^{i'j'kla'b'}(x) \nonumber\\ 
&&\hspace{11mm}
    +  K_{i'j'kl;a'b'c'd'}(x',x;t) 
    \overline{J}^{i'j'kla'b'c'd'}(x) \Bigr\} \ ,
\label{b7}
\end{eqnarray}
where
\begin{equation}
\begin{array}{lcl}
J^{i'j'kl} 
  &=&  \sqrt{h} 
       \Bigl\{ ( \frac{1}{4} h^{i'j'} h^{kl} 
               - \frac{1}{2} h^{i'k} h^{j'l} ) R^2 
  \vspace{1mm}\\
  & &~~~~~  + ( 2 h^{i'k} R^{j'l} + 2 h^{j'k} R^{i'l} 
                - h^{i'j'} R^{kl} -   h^{kl} R^{i'j'} ) 
                R \vspace{1mm}\\ 
  & &~~~~~  + 2 R^{i'j'} R^{kl}  
            +  ( 2 h^{ka'} h^{lb'} - 2 h^{kl} h^{a'b'} ) 
            {R^{i'j'}}_{;b'a'}  
            \vspace{1mm}\\
  & &~~~~~  +   (   h^{kl} h^{i'j'} h^{a'b'}
                - 2 h^{kl} h^{i'a'} h^{j'b'}       
                - 2 h^{i'k} h^{j'l} h^{a'b'}  \vspace{1mm}\\
  & & ~~~~~~    -   h^{ka'} h^{lb'} h^{i'j'} 
                + 2 h^{i'k} h^{la'} h^{j'b'}  
                + 2 h^{j'k} h^{lb'} h^{i'a'} ) 
                R_{;b'a'}   \Bigr\}  \ , 
     \vspace{3mm} \\
J^{i'j'kla'b'}  
  &=&  \sqrt{h}
     \Bigl\{   ( h^{i'j'} h^{kl} h^{a'b'}     
           + h^{i'k}  h^{j'b'} h^{la'} 
           + h^{b'i'} h^{j'l}  h^{ka'}  \vspace{1mm}\\
  & &~~~~~ - h^{i'k}  h^{j'l}  h^{a'b'}  
           - h^{i'j'} h^{kb'}  h^{la'} 
           - h^{kl}   h^{i'a'} h^{j'b'} ) R \vspace{1mm}\\ 
  & &~~~~ + ( 2 h^{ka'} h^{lb'} - 2 h^{kl} h^{a'b'}) R^{i'j'} 
          + ( 2 h^{i'a'} h^{j'b'} 
            - 2 h^{i'j'} h^{a'b'}) R^{kl}  \Bigr\}  \ ,
       \vspace{3mm} \\
J^{i'j'kla'b'c'd'}  
  &=& 2 \sqrt{h}    (   h^{i'j'} h^{a'b'} h^{kl} h^{c'd'} 
                - h^{i'a'} h^{j'b'} h^{kl} h^{c'd'} \vspace{1mm} \\
 & &~~~~~       - h^{i'j'} h^{a'b'} h^{kc'} h^{ld'}
                + h^{i'a'} h^{j'b'} h^{kc'} h^{ld'} )  \ , 
       \vspace{3mm}\\
\overline{J}^{i'j'kl}  &=&
    \sqrt{h} \Bigl\{ ( \frac{1}{4} h^{i'j'} h^{kl} 
         - \frac{1}{2} h^{i'k} h^{j'l} ) R_{a'b'} R^{a'b'} 
         +  2   R^{i'k} R^{j'l}  \\
 & &~~~~  - h^{kl} {R^{i'}}_{a'} R^{j'a'} 
          - h^{i'j'} {R^{k}}_{a'} R^{la'}
          + 4 h^{i'k} {R^{j'}}_{a'} R^{la'}  \\  
 & &~~~~  + (\frac{1}{2} h^{i'j'} h^{a'b'} 
          -   h^{i'a'} h^{j'b'} ) {R^{kl}}_{;b'a'}         
          + (  2 h^{i'a'} h^{kb'}  
          - 2 h^{i'k}  h^{a'b'} ) {R^{j'l}}_{;b'a'}   
 \vspace{1mm} \\
 & &~~~~  + (  2 h^{i'b'} h^{j'k} 
          -   h^{i'j'} h^{kb'}  ) {R^{la'}}_{;b'a'}        
          + (\frac{1}{2} h^{i'j'} h^{kl}   
          -   h^{i'k}  h^{j'l}  ) {R^{a'b'}}_{;b'a'}  
\vspace{1mm}\\       
 & &~~~~  + 2 h^{i'k} h^{lb'} {R^{j'a'}}_{;b'a'} 
          - 2 h^{a'i'} h^{kl} {R^{j'b'}}_{;b'a'} \Bigr\}  \ , 
\vspace{3mm}\\    
\overline{J}^{i'j'kla'b'}  &=&
    \sqrt{h} \Bigl\{   
            ( 2 h^{i'b'} h^{ka'} - 2 h^{a'b'} h^{i'k} ) R^{j'l} 
          - 2 h^{i'b'} h^{kl} R^{j'a'}  \vspace{1mm}\\
 & &~~~~  + 2 h^{i'l}  h^{ka'} R^{j'b'}  
          +  ( 2 h^{i'b'} h^{j'k} -  h^{i'j'} h^{kb'} ) R^{la'}
\vspace{1mm} \\
 & &~~~~  - h^{i'j'} h^{ka'} R^{lb'} 
          +  (   h^{i'j'} h^{kl} -   h^{i'k} h^{j'l} ) R^{a'b'} 
       \Bigr\}  \ ,\vspace{3mm} \\
\overline{J}^{i'j'kla'b'c'd'}  &=&
    \frac{1}{2} \sqrt{h} \Bigl\{ h^{a'b'} h^{i'k} h^{j'l} h^{c'd'} 
          -  h^{a'b'} h^{i'l} h^{j'd'} h^{kc'} 
          -  h^{a'b'} h^{i'd'} h^{j'l} h^{kc'} \vspace{1mm}\\
 & &~~~~~ +  h^{a'b'} h^{i'd'} h^{j'c'} h^{kl}   
          +  h^{i'j'} h^{ka'} h^{lb'} h^{c'd'} 
          -  h^{i'j'} h^{la'} h^{b'd'} h^{kc'} \vspace{1mm}\\
 & &~~~~~ -  h^{i'j'} h^{a'd'} h^{lb'} h^{kc'}   
          + \frac{1}{2} h^{i'j'} h^{a'd'} h^{b'c'} h^{kl}   
          + \frac{1}{2} h^{i'j'} h^{a'c'} h^{b'd'} h^{kl}  
                                               \vspace{1mm}\\
 & &~~~~~ -2 h^{i'b'} h^{ka'} h^{j'l} h^{c'd'} 
          +2 h^{i'b'} h^{la'} h^{j'd'} h^{kc'} 
          +2 h^{i'b'} h^{a'd'} h^{j'l} h^{kc'}  \vspace{1mm}\\
 & &~~~~~ -  h^{i'b'} h^{a'd'} h^{j'c'} h^{kl} 
          -  h^{i'b'} h^{a'c'} h^{j'd'} h^{kl}  \Bigr\} \ .
\end{array}
\label{b8}
\end{equation}
Here we have not written $J^{i'j'kla'}$ and 
$\overline{J}^{i'j'kla'}$ because they 
do not contribute to the final results (\ref{209}) 
and (\ref{210}). Substitution 
of (\ref{a9}) to (\ref{a13}) into (\ref{b6}) 
and (\ref{b7}) leads to 
\begin{eqnarray}
\lefteqn{\Delta(x;t)  \int d^3y \sqrt{h} R^2} \nonumber \\ 
&&= \frac{\sqrt{h(x)}}{(4 \pi t)^{3/2}} \Bigl\{~ 
    \frac{1}{t} \Bigl( - 3  R(x) \Bigr) 
                  + \Bigl( \frac{11}{8} - 3 ~ \xi \Bigr) R^2(x)  
   + \frac{31}{6} R_{ij}(x) R^{ij}(x)  
   - {R^{;i}}_i (x) ~  \Bigr\} \nonumber\\
&&~~~~ + O(t^{-\frac{1}{2}}) \ , 
\label{b9}  \\
\lefteqn{\Delta(x;t)  \int d^3y \sqrt{h} R_{ij} R^{ij} }
\nonumber \\
&&=  \frac{\sqrt{h(x)}}{(4 \pi t)^{3/2}} \Bigl\{~ 
     \frac{1}{t^2} \Bigl(\frac{15}{4}  \Bigr) 
   + \frac{1}{t} 
   \Bigl( - \frac{13}{8} + \frac{15}{4}~ \xi \Bigr) R(x)  
   + \Bigl( - \frac{241}{480} - \frac{13}{8} ~ \xi 
       + \frac{15}{8} ~ \xi^2 \Bigr) R^2 (x) \nonumber \\
&&\hspace{2cm} 
   + \frac{97}{20} R_{ij}(x) R^{ij}(x)   
   + \Bigl(- \frac{11}{30} - \frac{25}{24} ~ \xi \Bigr) 
   {R^{;i}}_i (x) ~\Bigr\} 
   + O(t^{-\frac{1}{2}}) \ .  
\label{b10}
\end{eqnarray}

%%%%%%%%%%%%%%%%%%%%%%%%%%%%%%%%%%%%%%%%%%%%%%%%%%%%%%%%%%%%%%%%
    
\section{{\normalsize\bf CONSISTENCY OF THE CONSTRAINTS}}

In this appendix we will compute the commutator (\ref{304}). 
Let us first consider 
the commutator 
\begin{equation}
\Bigl[~~ \int d^3x~ \eta_1(x) ~\Delta_R(x) ~~, 
~~ \int d^3y~ \eta_2(y)~ \Delta_R(y) ~~ \Bigr] \ ,
\label{c1}
\end{equation}
where $\eta_1$ and $\eta_2$ are arbitrary scalar functions. 
We note that (\ref{c1}) is antisymmetric 
under the exchange of $\eta_1$ and
$\eta_2$. The above commutator can be obtained from 
$[ \int d^3x \eta_1(x) \Delta(x;t),  
   \int d^3y \eta_2(x) \Delta(y;t') ]$ 
by analytic continuation with respect to $t$ and $t'$, 
as discussed in Sec. II.
\begin{eqnarray}
\lefteqn{
\Bigl[~ \displaystyle{\int} d^3x~ \eta_1(x) ~\Delta(x;t) ~~, 
~~ \int d^3y~ \eta_2(y)~ \Delta(y;t') ~ \Bigr] } \nonumber\\
&&= \displaystyle{\int} d^3x d^3x' d^3y d^3y' 
     ~\eta_1(x) ~\eta_2(y)~ K_{i'j'kl}(x',x;t) \nonumber\\
&&~~~\times \biggl\{~ 
   \biggl( \displaystyle{\frac{\delta}{\delta h_{i'j'}(x')}} 
        K_{a'b'cd}(y',y;t') \biggr) 
        \frac{\delta}{\delta h_{kl}(x)}  
 + \biggl( \displaystyle{\frac{\delta}{\delta h_{kl}(x)} } 
        K_{a'b'cd}(y',y;t') \biggr) 
        \frac{\delta}{\delta h_{i'j'}(x')} \nonumber\\
&&~~~~~~ 
 + \biggl( \displaystyle{ 
        \frac{\delta}{\delta h_{kl}(x)} 
        \frac{\delta}{\delta h_{i'j'}(x')}   
        K_{a'b'cd}(y',y;t') \biggr) ~\biggr\} 
        \frac{\delta}{\delta h_{cd}(y)}
        \frac{\delta}{\delta h_{a'b'}(y')} } \nonumber\\  
&&~-~ (~ \eta_1 ~\leftrightarrow ~\eta_2 ~) \ .
\label{c2}
\end{eqnarray}
In our renormalization prescription, the cutoff $t$ 
is removed by analytic 
continuation. This implies that we can take 
the naive limit $t \rightarrow 0$ 
if this limit produces no divergences 
in resulting calculations. Thus we 
may take the limit $t' \rightarrow 0$ 
in the first three terms on the right 
hand side of (\ref{c2}) and replace $K_{a'b'cd}(y',y;t')$ 
by $G_{a'b'cd}(y) \delta(y',y)$\footnote{ 
In this paper, we will not rigorously 
justify taking the limit $t' \rightarrow 0$ 
at this stage, which might produce 
potential divergences due to the second functional derivatives 
$ \frac{\delta}{\delta h_{cd}(y)} 
  \frac{\delta}{\delta h_{a'b'}(y)} $ 
at the same point $y$.}.
Then, the first three terms on the right hand 
side of (\ref{c2}) are reduced to 
\begin{eqnarray}
\lefteqn{
\displaystyle{\int} d^3x d^3y~ \eta_1(x)~\eta_2(y) }\nonumber\\
&&\times \biggl\{ ~ 
    K_{i'j'kl}(y,x;t) g^{i'j'}_{a'b'cd}(y) 
     \displaystyle{\frac{\delta}{\delta h_{kl}(x)}}   
   + \displaystyle{\int} d^3x' 
     K_{i'j'kl}(x',x;t) g^{kl}_{a'b'cd}(y) \delta(y,x) 
     \displaystyle{
       \frac{\delta}{\delta h_{i'j'}(x')} } \nonumber\\
&&~~~ + K_{i'j'kl}(y,x;t) 
        g^{i'j'kl}_{a'b'cd}(y)~ \delta(y,x) ~ \biggr\} 
\displaystyle{ 
 \frac{\delta}{\delta h_{cd}(y)} 
 \frac{\delta}{\delta h_{a'b'}(y)} } \ ,
\label{c3}
\end{eqnarray}
where
\begin{equation}
\begin{array}{l}
\displaystyle{
  \frac{\delta}{\delta h_{i'j'}(x')} } G_{a'b'cd}(y) 
  \equiv g^{i'j'}_{a'b'cd}(y)~ \delta(y,x')  \ , \vspace{2mm}\\
\displaystyle{
\frac{\delta}{\delta h_{kl}(x)} 
\frac{\delta}{\delta h_{i'j'}(x')} } G_{a'b'cd}(y) 
 \equiv g^{i'j'kl}_{a'b'cd}(y)~ \delta(y,x') ~\delta(y,x) \ .
\end{array}
\label{c4}
\end{equation}
Taking the limit $t \rightarrow 0$ 
in the first two terms of (\ref{c3}) but 
keeping $t$ nonzero in the last term of (\ref{c3}), we have 
\begin{eqnarray}
\lefteqn{\displaystyle{\int} d^3x~ \eta_1(x) ~ \eta_2(x)~  
\biggl\{~ 
    G_{i'j'kl}(x)~ g^{i'j'}_{a'b'cd}(x) 
         \displaystyle{\frac{\delta}{\delta h_{kl}(x)}} } 
         \nonumber\\
&&+~ G_{i'j'kl}(x)~ g^{kl}_{a'b'cd}(x) 
         \displaystyle{\frac{\delta}{\delta h_{i'j'}(x)}} 
  + K_{i'j'kl}(x,x;t)~ g^{i'j'kl}_{a'b'cd}(x) ~ \biggr\} 
         \displaystyle{
         \frac{\delta}{\delta h_{cd}(x)}
         \frac{\delta}{\delta h_{a'b'}(x)} } \ ~.
 \nonumber \\
\label{c5}
\end{eqnarray} 
This is clearly symmetric 
under the exchange of $\eta_1$ and $\eta_2$. 
Since only antisymmetric terms under 
the exchange of $\eta_1$ and $\eta_2$ 
survive in the commutator (\ref{c1}), we conclude that 
\begin{equation}
\Bigl[~~ \int d^3x~ \eta_1(x) ~\Delta_R(x) ~~, 
~~ \int d^3y~ \eta_2(y)~ \Delta_R(y) ~~ \Bigr] ~ = ~ 0 ~\ .
\label{c6}
\end{equation}
Therefore, nontrivial contributions 
to the commutator (\ref{304}) come from 
\begin{eqnarray}
&&\Bigl[~~ - 16 \pi G \displaystyle{\int} 
d^3x~ \eta_1(x) ~ \Delta_R(x) ~~, ~~
\displaystyle{\frac{1}{16 \pi G} \int} d^3y~ \eta_2(y)~ 
         \sqrt{h(y)}~ (R(y) + 2 \Lambda)~  \Bigr]   \nonumber \\
&&~~~~~~-~ (~ \eta_1 ~\leftrightarrow ~\eta_2 ~) \ .
\label{c7}
\end{eqnarray}
As before, the above commutator can be obtained, 
by analytic continuation 
with respect to $t$, from 
\begin{eqnarray}
\lefteqn{
\Bigl[~~ - 16 \pi G 
\displaystyle{\int} d^3x~ \eta_1(x) ~ \Delta(x;t)  ~~, ~~
\displaystyle{\frac{1}{16 \pi G} \int} d^3y~ \eta_2(y)~ 
         \sqrt{h(y)}~ (R(y) + 2 \Lambda)~  \Bigr] }\nonumber\\ 
&&~-~ (~ \eta_1 ~\leftrightarrow ~\eta_2 ~)   \nonumber\\
\lefteqn{= \displaystyle{\int} d^3x d^3x' d^3y  
     ~\eta_1(x) ~\eta_2(y)~ K_{i'j'kl}(x',x;t) }\nonumber\\
&&\times \biggl\{~ 
 - \biggl( \displaystyle{\frac{\delta}{\delta h_{i'j'}(x')}} 
           \sqrt{h(y)}~ (R(y) + 2 \Lambda) \biggr) 
           \frac{\delta}{\delta h_{kl}(x)} 
- \biggl( \displaystyle{\frac{\delta}{\delta h_{kl}(x)} } 
           \sqrt{h(y)}~ (R(y) + 2 \Lambda)  \biggr) 
           \frac{\delta}{\delta h_{i'j'}(x')} \nonumber\\
&&~~~ 
 - \biggl( \displaystyle{ 
           \frac{\delta}{\delta h_{kl}(x)}
           \frac{\delta}{\delta h_{i'j'}(x')} } 
           \sqrt{h(y)}~ (R(y) + 2 \Lambda) \biggr) 
           ~\biggr\} \nonumber\\
&&-~ (~ \eta_1 ~\leftrightarrow ~\eta_2 ~) \  .
\label{c8}
\end{eqnarray}
We can take the limit $t \rightarrow 0$ 
without any trouble in the first two 
terms on the right hand side of (\ref{c8}). 
It is then easy to see that these 
two terms give the first term 
on the right hand side of (\ref{304}). 
We cannot, however, take the limit $t \rightarrow 0$ 
in the third term on 
the right hand side of (\ref{c8}). 
The third term would give an anomalous 
contribution to the commutator. 
It is not difficult to show that 
\begin{equation}
\begin{array}{l} 
\displaystyle{\int} 
d^3x d^3x' d^3y ~ \eta_1(x)~ \eta_2(y) ~K_{i'j'kl}(x',x;t)
\biggl( - \displaystyle{ 
        \frac{\delta}{\delta h_{kl}(x)}
        \frac{\delta}{\delta h_{i'j'}(x')} } \biggr)  
 \sqrt{h(y)}~ (R(y) + 2 \Lambda) \\
 ~~~~~~~~~-~ 
 (~ \eta_1 ~\leftrightarrow ~\eta_2 ~)  \vspace{4mm}\\ 
=  \displaystyle{ 
- \frac{\frac{1}{24} + \xi}{(4 \pi)^{3/2} t^{1/2}} }
  \displaystyle{\int} d^3x 
  \sqrt{h(x)} ~\Bigl(  \eta_1(x) ( \nabla\!_i\, \eta_2(x) ) 
  - (\nabla\!_i\, \eta_1 (x)) \eta_2 (x) \Bigr)~
   \nabla^i\! R(x)   + O(t^{\frac{1}{2}}) \ .
\end{array}
\label{c9}
\end{equation}
Therefore, our renormalization prescription tells us that 
\begin{eqnarray}
&&\Bigl[~~ - 16 \pi G 
\displaystyle{\int} d^3x~ \eta_1(x) ~ \Delta_R(x) ~~, ~~
\displaystyle{\frac{1}{16 \pi G} \int} d^3y~ \eta_2(y)~ 
         \sqrt{h(y)}~ (R(y) + 2 \Lambda)~  \Bigr] \nonumber \\
&&~~~~~~-~ (~ \eta_1 ~\leftrightarrow ~\eta_2 ~)  \nonumber\\
&&= \displaystyle{\int} d^3x ~ 
\Bigl(  \eta_1(x) ( \nabla\!_i\, \eta_2(x) ) 
  - (\nabla\!_i\, \eta_1 (x)) \eta_2 (x) \Bigr) \nonumber \\
&&\hspace{3cm}\times \biggl(~ i ~{\cal H}^i(x)
- \displaystyle{\frac{\frac{1}{24} + \xi}{(4 \pi)^{3/2}} }
\phi^{(1)}(0) \sqrt{h(x)}~ \nabla^i\! R(x)~ \biggr)   \ .  
\label{c10}
\end{eqnarray}
Combining (\ref{c6}) with (\ref{c10}), 
we get the result (\ref{304}) with 
the anomalous term (\ref{305}).
  
%%%%%%%%%%%%%%%%%%%%%%%%%%%%%%%%%%%%%%%%%%%%%%%%%%%%%%%%%%%%%%%%%%


\newpage

\begin{thebibliography}{99}
\baselineskip = 6.5mm
\bibitem{nonrenormalizability}G. 't Hooft and M. Veltman,  
    Ann. Inst. Henri Poincar\'e 
    \underline{20}, 69 (1974) ; 
    M. H. Goroff and A. Sagnotti, Nucl. Phys. 
    \underline{B266}, 709 (1986).  

\bibitem{minimum length}
L. J. Garay, Int. J. Mod. Phys. A \underline{10}, 145 (1995). 

\bibitem{discrete spacetime}
P. Gibbs, \lq\lq The Small Scale Structure of Space-Time", 
     PEG-06-95, hep-th/9506171.

\bibitem{TQFT}
E. Witten, Commun. Math. Phys. \underline{117}, 353 (1988) ; 
\underline{118}, 411 (1988) ; 
Phys. Lett. B \underline{206}, 601 (1988). 

\bibitem{dimensional reduction}G. 't Hooft, 
\lq\lq Dimensional reduction in quantum gravity", 
THU-93/26, gr-qc/9310026.

\bibitem{hologram}
L. Susskind, J. Math. Phys. \underline{36}, 6377 (1995).

\bibitem{smolin}L. Smolin, 
\lq\lq The Bekenstein bound, topological quantum field theory and
pluralistic quantum cosmology", CGPG-9518-7, gr-qc/9508064 ; 
J. Math. Phys. \underline{36}, 6417 (1995). 

\bibitem{Ashtekar}
A. Ashtekar, Phys. Rev. Lett. \underline{57}, 2244 (1986).

\bibitem{loop}
T. Jacobson and L. Smolin, Nucl. Phys.
 \underline{B299}, 295 (1988) ; 
C. Rovelli and L. Smolin, Nucl. Phys. 
\underline{B331}, 80 (1990). 

\bibitem{area operator}
A. Ashtekar, C. Rovelli and L. Smolin, Phys. Rev. Lett. 
\underline{69}, 237 (1992).

\bibitem{volume}
C. Rovelli and L. Smolin, 
Nucl. Phys. \underline{B442}, 593 (1995).

\bibitem{WDW}J. A. Wheeler, in {\it Battelle Rencontres}, edited by 
C. DeWitt and J. A. Wheeler (Benjamin, New York, 1968) ; 
B. S. DeWitt, Phys. Rev. \underline{160}, 1113 (1967).

\bibitem{analysis}T. Horiguchi, K. Maeda and M. Sakamoto, 
Phys. Lett. B
\underline{344}, 105 (1995).

\bibitem{Mansfield}
P. Mansfield, Nucl. Phys. \underline{B418}, 113 (1994).

\bibitem{light-front QCD}
K. G. Wilson, et al, Phys. Rev. D \underline{49}, 6720 (1994). 

\bibitem{Schwinger-DeWitt}
J. S. Schwinger, Phys. Rev. \underline{82}, 664 (1951) ;
B. S. DeWitt, Phys. Rep. D \underline{42}, 2548 (1975). 

\bibitem{dimred}
G. 't Hooft and M. Veltman, 
Nucl. Phys. \underline{B44}, 189 (1972).

\bibitem{strong}
M. Henneaux, M. Pilati and C. Teitelboim, 
Phys. Lett. B \underline{110}, 123 (1982) ;
C. Teitelboim, Phys. Rev. D \underline{25}, 3159 (1982) ; 
M. Pilati, Phys. Rev. D \underline{26}, 2645 (1982) ; 
C.Rovelli, Phys. Rev. D \underline{35}, 2987 (1987) ; 
A. Ashtekar, 
{\it New Perspectives in Canonical Gravity} 
(Bibliopolis, Napoli, 1988). 

\bibitem{Kodama}
H. Kodama, Phys. Rev. D \underline{42}, 2548 (1990). 

\bibitem{3-d gravity}
E. Witten, Nucl. Phys. \underline{B311}, 46 (1988/89). 

\bibitem{Kowalski}
J. Kowalski-Glikman and K. A. Meissner, 
\lq\lq A Class of Exact Solutions of the 
Wheeler-DeWitt Equation", IFT/2/96, hep-th/9601062.

\bibitem{dilaton}
M. B. Green, J. H. Schwarz and E. Witten, {\it Superstring theory},
(Cambridge University Press, Cambridge, 1987); 
C. G. Callan, S. B. Giddings, J. A. Harvey and A. Strominger, 
Phys. Rev. 
D \underline{45}, 1005 (1992);
R. H. Brandenberger, \lq\lq Physics of the very early universe", 
BROWN-HET-1010, astro-ph/9509154  . 

\bibitem{dynamical theory}
B. S. DeWitt, {\it Dynamical Theory of Groups and Fields},
(Gordon and Breach, New York, 1965).

\end{thebibliography}
\end{document}